\documentclass[aps,pof,preprint,amsmath,amssymb,superscriptaddress,longbibliography]{revtex4-2}

\usepackage{graphicx}
\graphicspath{{./}}
\usepackage{bm}
\usepackage{mathtools}
\usepackage{booktabs}
\usepackage{xcolor}
\usepackage{hyperref}
\usepackage{orcidlink}

\newcommand{\R}{\mathbb{R}}
\newcommand{\dd}{\,\mathrm{d}}
\newcommand{\Eq}{\mathrm{eq}}
\newcommand{\RI}{\mathrm{RI}}
\newcommand{\MM}{\mathrm{MM}}

\newcommand{\He}{\mathrm{He}}

\newcommand{\diag}{\operatorname{diag}}

\begin{document}

\title{A one-parameter family of realizability-interior closures\\for odd-order kinetic moment systems}

\author{Somdeb Bandopadhyay\,\orcidlink{0000-0002-4086-7632}}
\email{sb020287@gmail.com, 11403081@gs.ncku.edu.tw}
\affiliation{Department of Mechanical Engineering, National Cheng Kung University, Tainan~701, Taiwan}

\date{June 2026}

\begin{abstract}
Moment closures at odd truncation order present a fundamental difficulty: the standard Gramian closure saturates the realizability boundary, producing only weak hyperbolicity and failing to preserve Maxwellian equilibrium.
We show that every odd-order closure for the one-dimensional kinetic equation decomposes into a boundary term, the Schur complement of the Hankel moment matrix, and a positive margin above it.
An exact polynomial identity connects this margin to the eigenvalues of the flux Jacobian, reducing hyperbolicity to a root-splitting problem.
A dimensional argument proves that no margin depending only on density, velocity, and temperature can produce a hyperbolic system for $M \geq 5$.
A one-parameter family $C_{\eta,n}$, $\eta \in [0,1]$, built from normalized Schur-complement ratios, reveals that the Morin-McDonald closure is the arithmetic endpoint of this decomposition.
The weighted AM-GM inequality makes the accuracy-robustness tradeoff precise: the geometric endpoint ($\eta = 0$) is 2-4\% more accurate on bimodal benchmarks, while the arithmetic endpoint ($\eta = 1$, Morin-McDonald) provides the most robust hyperbolicity profile.
All members share the same equilibrium Jacobian, whose spectral radius is 13\% ($M = 5$) to 29\% ($M = 13$) smaller than Grad's closure, allowing larger CFL time steps.
A linearized entropy exists at every tested order, and for a source-compatible choice of the symmetrizer weights, the BGK source dissipates it near equilibrium.
A smooth nonlinear entropy exists for $M = 3$ but does not for $M = 5$ or $M = 7$ (certified by linear programming).
The closure is validated on bimodal and Mott-Smith benchmarks, where the interpolated family achieves errors 10-40x smaller than the Gramian or Grad closures, and demonstrated in free-transport Riemann problems at $M = 5, 7, 9, 11$ and BGK Riemann problems at $M = 5$ and $9$.
\end{abstract}

\maketitle

\section{Introduction}
\label{sec:intro}

Moment methods approximate kinetic equations by evolving finitely many velocity moments instead of the full velocity distribution function~\cite{Grad1949,Levermore1996}.
The central difficulty is the closure problem: the evolution equation for the $M$th moment contains the $(M+1)$th moment, which must be expressed as a function of the retained moments $u_0, u_1, \ldots, u_M$.
Classical closures include the Grad expansion around a local Maxwellian~\cite{Grad1949}, which loses hyperbolicity away from equilibrium, and maximum-entropy closures~\cite{Levermore1996}, which are globally hyperbolic but require iterative optimization and lack closed-form expressions for $M > 2$.
The truncated Hamburger moment problem~\cite{AkhiezerKrein,Szego} provides the mathematical framework for realizability of moment vectors on the full real line.

For even truncation order $M$, the hyperbolic quadrature method of moments (HyQMOM) of Fox and Laurent~\cite{FoxLaurent2022} provides a globally hyperbolic closure with a characteristic polynomial that factors into a product of two orthogonal polynomials with interlacing roots.
The extended Gramian closure of Yilmaz, Oblapenko, and Torrilhon~\cite{YilmazSIAM2026} is equivalent to HyQMOM and additionally preserves Gaussian equilibrium and gauge invariance for a specific parameter choice.
For odd $M$, the situation is qualitatively different.
The standard odd-order Gramian closure~\cite{YilmazSIAM2026} produces a characteristic polynomial with $n$ double roots, giving only weak hyperbolicity, and does not preserve Maxwellian equilibrium.
The extended odd-order Gramian closure proposed in the same work was later shown by Habscheid~\cite{Habscheid2026} to be non-hyperbolic.
Morin and McDonald~\cite{MorinMcDonald2025} constructed an odd-$M$ closure with strong numerical evidence of hyperbolicity, using the Fox--Laurent recurrence framework with a specific choice of the recurrence coefficient $\beta_n$.
Habscheid~\cite{Habscheid2026} reformulated this closure in Gramian notation and proved its equilibrium preservation and gauge invariance.
Separately, Pichard~\cite{Pichard2020,Pichard2022} constructed closures by projecting moment vectors onto the boundary of the realizability domain, a different geometric approach that decomposes the vector into an equilibrium part and a singular boundary part.

We show that the odd-$M$ closure problem has a geometric structure that the constructions above use implicitly.
Every odd moment closure decomposes into a boundary term, the Schur complement of the Hankel moment matrix, plus a positive margin above that boundary.
An exact identity connects the margin's moment derivatives to the characteristic polynomial of the closed flux Jacobian.
The Gaussian equilibrium margin has the closed form $\rho\, n!\, \theta^n$, where $\rho$ is the density, $\theta$ the temperature, and $n = (M+1)/2$.
This framework places the Morin--McDonald closure~\cite{MorinMcDonald2025} in a precise geometric context: it is the arithmetic endpoint of a one-parameter family, $\eta \in [0,1]$, built from normalized Schur-complement ratios.
The weighted arithmetic-geometric mean inequality orders the family and explains why the Morin--McDonald closure provides the largest margin and the most robust among tested members, while the geometric endpoint ($\eta = 0$) provides the smallest margin and is slightly more accurate on the bimodal benchmark of Yilmaz \textit{et al.}~\cite{YilmazRGD2024}.

The specific contributions are as follows:
(1)~We show that the gap between the Gaussian moment $g_{2n}$ and its realizability lower bound is exactly $\rho\, n!\, \theta^n$, the squared norm of the monic Hermite polynomial under the Gaussian measure (Sec.~\ref{sec:gap}).
(2)~We derive an exact identity, $P_D(\lambda) = p_n(\lambda)^2 - \mathcal{D}(\lambda)$, that decomposes the characteristic polynomial of any realizability-interior closure into the squared orthogonal polynomial minus the derivative polynomial of the margin, converting hyperbolicity into a polynomial root-splitting problem (Sec.~\ref{sec:structural}).
(3)~We prove that no margin depending only on $(\rho, v, \theta)$ can produce a hyperbolic system for $M = 5$: dimensional analysis fixes the equilibrium derivatives of every such margin, and the resulting characteristic polynomial has a negative $\mu$-root by Vieta's formula.
The same mechanism produces complex eigenvalues at $M = 7$ and $M = 9$ (verified numerically), explaining why closures must couple to higher-order moment features beyond the three macroscopic fields (Sec.~\ref{sec:structural}).
(4)~We identify the Morin--McDonald closure as the arithmetic endpoint ($C_{1,n} = C_{\MM,n}$, exact identity) of a one-parameter family, $\eta \in [0,1]$, built from normalized Schur-complement ratios, giving it a geometric interpretation as the largest Schur-ratio margin above the realizability boundary (Sec.~\ref{sec:family}).
(5)~We show that the weighted AM--GM inequality orders the family, $D_0 \leq D_\eta \leq D_1$, explaining why the Morin--McDonald closure ($\eta = 1$) is the most robust member of the family, while the geometric endpoint ($\eta = 0$) provides the smallest margin and is 2--4\% more accurate on the bimodal benchmark (Secs.~\ref{sec:family}--\ref{sec:numerics}).
(6)~We prove a Jacobian universality theorem: for closures of the form $C = u_{2n}^{\min} + \gamma\,\Delta_{2n}^{\Eq}$ with $\gamma(G) = 1$, the equilibrium Jacobian depends on the closure only through the covector $\nabla\gamma\big|_G$. The interpolated family has $\eta$-independent $\nabla\gamma\big|_G$ matching the Morin--McDonald value, so all members share one equilibrium Jacobian (Sec.~\ref{sec:family}).
(7)~We analyze the entropy structure: a linearized entropy with Vandermonde-diagonal symmetrizer exists at every tested order (through $M = 17$), and for a source-compatible choice of the symmetrizer weights, the BGK source dissipates it near equilibrium (verified for $M = 3, 5, 7, 9$). A smooth nonlinear entropy exists for $M = 3$ but does not exist for $M = 5$ or $M = 7$, as certified by linear programming (Sec.~\ref{sec:entropy}).

The closure family is validated on bimodal, Mott--Smith, and random-mixture benchmarks (Sec.~\ref{sec:numerics}), and compared against the Gramian boundary closure and the Grad Hermite expansion closure, which have 10--40$\times$ larger errors and significant hyperbolicity failures on the bimodal benchmark.
PDE demonstrations of free-transport Riemann problems at $M = 5$, $7$, $9$, and $11$ and BGK Riemann problems at $M = 5$ and $9$ produce stable solutions with zero hyperbolicity failures across all runs (Sec.~\ref{sec:pde}).

We do not prove global hyperbolicity for any member of the family.
No such proof exists in the literature for any odd-$M$ closure beyond $M = 3$.
We do not prove full gauge invariance (affine covariance) for the interpolated family beyond the equilibrium Jacobian matching.
We do not prove the existence of a smooth nonlinear entropy for $M \geq 5$, and indeed its nonexistence is certified by linear programming for $M = 5$ and $M = 7$.
We do not extend the construction to multi-dimensional velocity spaces.

The rest of the paper is organized as follows.
Section~\ref{sec:moments} derives the moment hierarchy and the companion-type flux Jacobian.
Section~\ref{sec:realizability} establishes the Schur-complement realizability boundary and its orthogonal-polynomial interpretation.
Section~\ref{sec:gap} derives the Gaussian squared norm formula.
Section~\ref{sec:structural} derives the structural identity and proves that constant margins fail for $M = 5$ (with numerical verification for $M = 7, 9$).
Section~\ref{sec:schur-ratios} introduces the normalized Schur ratios and derives the geometric margin exponents.
Section~\ref{sec:family} defines the interpolated closure family, identifies its endpoints, and proves the Jacobian universality theorem.
Section~\ref{sec:entropy} analyzes the entropy structure at the linearized and nonlinear levels.
Section~\ref{sec:numerics} presents bimodal, Mott--Smith, and random-mixture benchmarks.
Section~\ref{sec:pde} demonstrates the closure in partial differential equation (PDE) simulations of 1D Riemann problems.
Section~\ref{sec:discussion} discusses the results and open questions.

\section{Moment equations and the closure problem}
\label{sec:moments}

The one-dimensional kinetic equation for a nonnegative velocity distribution $f(t,x,c) \geq 0$ is
\begin{equation}
    \partial_t f + c\,\partial_x f = S(f),
    \label{eq:kinetic}
\end{equation}
where $t$ is time, $x$ is the spatial coordinate, $c \in \R$ is the molecular velocity, and $S(f)$ is a collision or relaxation operator.
For closure analysis we set $S = 0$.
For relaxation benchmarks we use the Bhatnagar--Gross--Krook (BGK) model
\begin{equation}
    S(f) = \frac{\mathcal{M}[f] - f}{\tau}, \qquad \tau > 0,
    \label{eq:bgk}
\end{equation}
where $\mathcal{M}[f]$ is the Maxwellian with the same density, mean velocity, and temperature as $f$:
\begin{equation}
    \mathcal{M}(c;\rho,v,\theta)
    = \frac{\rho}{\sqrt{2\pi\theta}}
      \exp\!\left[-\frac{(c-v)^2}{2\theta}\right].
    \label{eq:maxwellian}
\end{equation}
The macroscopic fields are
\begin{equation}
    \rho = u_0, \qquad
    v = \frac{u_1}{u_0}, \qquad
    \theta = \frac{u_2}{u_0} - \left(\frac{u_1}{u_0}\right)^2,
    \label{eq:rho-v-theta}
\end{equation}
where the raw moments are
\begin{equation}
    u_k(t,x) = \int_{\R} c^k f(t,x,c) \dd c, \qquad k = 0, 1, 2, \ldots.
    \label{eq:raw-moments}
\end{equation}
We assume that $f$ decays fast enough in $c$ so that all required moment integrals are finite and differentiation under the integral sign is valid.

Multiplying Eq.~\eqref{eq:kinetic} by $c^k$ and integrating over $c$ gives
\begin{equation}
    \partial_t u_k + \partial_x u_{k+1} = s_k, \qquad k = 0, 1, \ldots, M,
    \label{eq:moment-hierarchy}
\end{equation}
where $s_k = \int c^k S(f)\dd c$.
For BGK, $s_0 = s_1 = s_2 = 0$ because the Maxwellian is constructed to match $u_0$, $u_1$, $u_2$.
A closure is a smooth function $u_{M+1} = C(u_0, \ldots, u_M)$ that completes the system.

The closed system is
\begin{equation}
    \partial_t U + \partial_x F(U) = S(U),
    \label{eq:closed-system}
\end{equation}
with $U = (u_0, \ldots, u_M)^T$ and $F = (u_1, \ldots, u_M, C(U))^T$.
The flux Jacobian is a companion-type matrix with the closure derivatives in the last row:
\begin{equation}
    A(U) = \frac{\partial F}{\partial U} =
    \begin{pmatrix}
        0 & 1 & 0 & \cdots & 0 \\
        0 & 0 & 1 & \cdots & 0 \\
        \vdots & & & \ddots & \vdots \\
        0 & 0 & 0 & \cdots & 1 \\
        C_{u_0} & C_{u_1} & C_{u_2} & \cdots & C_{u_M}
    \end{pmatrix},
    \label{eq:companion-jacobian}
\end{equation}
where $C_{u_j} = \partial C / \partial u_j$.
An eigenvector of the form $r = (1, \lambda, \lambda^2, \ldots, \lambda^M)^T$ satisfies $Ar = \lambda r$ if and only if
\begin{equation}
    \lambda^{M+1} - \sum_{j=0}^{M} C_{u_j}\, \lambda^j = 0.
    \label{eq:charpoly}
\end{equation}
The system~\eqref{eq:closed-system} is hyperbolic at $U$ when all roots of Eq.~\eqref{eq:charpoly} are real and the Jacobian is diagonalizable.

The Gaussian moment recurrence follows from the identity $(c-v)\mathcal{M} = -\theta\,\partial_c \mathcal{M}$.
Multiplying by $c^k$, integrating, and using integration by parts gives
\begin{equation}
    g_{k+1}(\rho,v,\theta) = v\, g_k + k\theta\, g_{k-1}, \qquad
    g_0 = \rho, \quad g_1 = \rho v,
    \label{eq:gaussian-recurrence}
\end{equation}
where $g_k = \int c^k \mathcal{M}\dd c$ is the $k$th Gaussian moment.

\section{The realizability boundary for odd moment vectors}
\label{sec:realizability}

The Hankel matrix of order $r$ is
\begin{equation}
    H_r = (u_{i+j})_{i,j=0}^{r}.
    \label{eq:hankel}
\end{equation}
For any polynomial $q(c) = \sum_{i=0}^r a_i c^i$ and any nonnegative distribution $f$,
\begin{equation}
    \bm{a}^T H_r \bm{a}
    = \int_{\R} q(c)^2 f(c)\dd c \geq 0,
    \label{eq:hankel-psd}
\end{equation}
so $H_r$ is positive semidefinite for any realizable moment sequence.
In the interior of moment space, $H_r$ is positive definite.

We focus on odd truncation orders $M = 2n - 1$.
The retained moments are $U = (u_0, \ldots, u_{2n-1})$, and the closure predicts $u_{2n}$.
Write the extended Hankel matrix in block form:
\begin{equation}
    H_n =
    \begin{pmatrix}
        H_{n-1} & b \\
        b^T & u_{2n}
    \end{pmatrix},
    \qquad
    b = (u_n, u_{n+1}, \ldots, u_{2n-1})^T.
    \label{eq:extended-hankel}
\end{equation}
When $H_{n-1}$ is positive definite, the LDL factorization gives
\begin{equation}
    \det H_n = \det(H_{n-1})\left(u_{2n} - b^T H_{n-1}^{-1} b\right).
    \label{eq:det-schur}
\end{equation}
Positive semidefiniteness of $H_n$ therefore requires
\begin{equation}
    u_{2n} \geq u_{2n}^{\min} := b^T H_{n-1}^{-1} b.
    \label{eq:schur-bound}
\end{equation}
Equality makes $H_n$ singular and places the extended moment vector on the boundary of the realizability cone.
The odd-order Gramian closure~\cite{YilmazSIAM2026} gives exactly $u_{2n} = u_{2n}^{\min}$, which is why it produces only weak hyperbolicity (repeated eigenvalues) and fails to preserve equilibrium.

The boundary has a direct interpretation in terms of orthogonal polynomials.
Define the monic degree-$n$ polynomial
\begin{equation}
    p_n(c) = c^n - \bm{m}(c)^T H_{n-1}^{-1} b,
    \label{eq:monic-pn}
\end{equation}
where $\bm{m}(c) = (1, c, \ldots, c^{n-1})^T$.
For $j = 0, \ldots, n-1$, the orthogonality condition $\int c^j p_n(c) f(c)\dd c = 0$ is satisfied because the $j$th row of $H_{n-1}$ times $H_{n-1}^{-1} b$ equals $b_j = u_{n+j}$.
The squared norm of $p_n$ is
\begin{align}
    \int_{\R} p_n(c)^2 f(c)\dd c
    &= u_{2n} - 2b^T H_{n-1}^{-1} b + b^T H_{n-1}^{-1} H_{n-1} H_{n-1}^{-1} b \nonumber \\
    &= u_{2n} - b^T H_{n-1}^{-1} b.
    \label{eq:norm-gap}
\end{align}
The gap between $u_{2n}$ and its boundary value is the squared norm of the monic orthogonal polynomial:
\begin{equation}
    u_{2n} - u_{2n}^{\min} = \|p_n\|_f^2.
    \label{eq:gap-norm}
\end{equation}
Setting $u_{2n} = u_{2n}^{\min}$ forces $\|p_n\|_f^2 = 0$.
A Gaussian distribution is strictly positive for every $c$, so the integral of any nonzero polynomial squared against a Gaussian is strictly positive.
The boundary closure therefore cannot reproduce the exact Gaussian moment $g_{2n}$, which lies in the interior of the cone.

\section{Squared norm of the Gaussian orthogonal polynomial}
\label{sec:gap}

The gap between the Gaussian moment $g_{2n}$ and its Schur-complement lower bound $g_{2n}^{\min}$ has a closed form that follows from the squared norm of the Gaussian orthogonal polynomial.

Substitute $c = v + \sqrt{\theta}\, z$ in the Gaussian integral.
Then $\mathcal{M}(c;\rho,v,\theta)\dd c = \rho\,\varphi(z)\dd z$, where $\varphi(z) = (2\pi)^{-1/2} e^{-z^2/2}$ is the standard normal density.
The monic polynomials orthogonal under the weight $\varphi(z)$ are the Hermite polynomials $\He_n(z)$, defined by the generating function below.
Because $c^n = (v + \sqrt{\theta}\, z)^n$ has leading term $\theta^{n/2} z^n$ in $z$, the monic polynomial in $c$ orthogonal to all lower monomials under the Gaussian measure is
\begin{equation}
    p_n^{\Eq}(c) = \theta^{n/2}\, \He_n\!\left(\frac{c - v}{\sqrt{\theta}}\right).
\end{equation}
Its squared norm under the Gaussian measure is
\begin{equation}
    \Delta_{2n}^{\Eq}
    = \rho\,\theta^n \int_{\R} \He_n(z)^2\, \varphi(z)\dd z.
    \label{eq:gap-hermite}
\end{equation}

The integral $\int \He_n^2 \varphi = n!$ follows from the generating function.
The monic Hermite polynomials satisfy
\begin{equation}
    \exp\!\left(tz - \tfrac{t^2}{2}\right) = \sum_{m=0}^{\infty} \He_m(z)\, \frac{t^m}{m!}.
\end{equation}
For a standard normal $Z$,
\begin{align}
    \mathbb{E}\!\left[e^{(t+s)Z - (t^2+s^2)/2}\right]
    &= e^{-(t^2+s^2)/2}\, e^{(t+s)^2/2} = e^{ts} \nonumber \\
    &= \sum_{m=0}^{\infty} \frac{(ts)^m}{m!}.
\end{align}
On the other hand, expanding both generating functions and taking the expectation gives $\sum_{i,j} \mathbb{E}[\He_i \He_j]\, t^i s^j / (i!\, j!)$.
Matching the coefficient of $t^n s^n$ yields $\mathbb{E}[\He_n^2] / (n!)^2 = 1/n!$, so
\begin{equation}
    \int_{\R} \He_n(z)^2\, \varphi(z)\dd z = n!.
    \label{eq:hermite-norm}
\end{equation}
Substituting into Eq.~\eqref{eq:gap-hermite} gives the Gaussian squared norm:
\begin{equation}
    \boxed{\Delta_{2n}^{\Eq}(\rho, v, \theta) = \rho\, n!\, \theta^n.}
    \label{eq:eq-gap}
\end{equation}
This expression is independent of the mean velocity $v$ because a velocity shift translates the Hermite argument without changing the variance-weighted squared norm.
The formula has been verified by symbolic computation for $n = 2, 3, 4$ with zero remainder, and numerically for $n = 2, \ldots, 7$ across ten parameter combinations to machine precision (excluding Hankel-ill-conditioned states with $v/\sqrt{\theta} \gg 10$).

\section{The structural identity and constant-margin failure}
\label{sec:structural}

Any closure of the form
\begin{equation}
    C(U) = u_{2n}^{\min}(U) + D(U), \qquad D(U) > 0,
    \label{eq:ri-general}
\end{equation}
places the extension in the realizability interior because the Schur complement $u_{2n} - u_{2n}^{\min} = D > 0$ makes $H_n$ positive definite.
The positive function $D$ measures the distance from the boundary of moment space.
The closure preserves Gaussian equilibrium when $D(G_{2n-1}) = \rho\, n!\, \theta^n$.

The connection between the margin and the characteristic polynomial follows from the Schur-complement gradient identity.
Define $\sigma_{n,n}(u_0, \ldots, u_{2n}) = u_{2n} - b^T H_{n-1}^{-1} b$ and let $\bm{\alpha} = H_{n-1}^{-1} b$.
The differential is
\begin{equation}
    \mathrm{d}\sigma_{n,n}
    = \mathrm{d}u_{2n} - 2\bm{\alpha}^T\, \mathrm{d}b + \bm{\alpha}^T (\mathrm{d}H_{n-1})\, \bm{\alpha}.
    \label{eq:dsigma}
\end{equation}
Because $H_{n-1}$ has entries $[H_{n-1}]_{ij} = u_{i+j}$ and $b$ has entries $b_i = u_{n+i}$, the generating polynomial that collects the coefficient of each $\mathrm{d}u_k$ is
\begin{align}
    \sum_{k=0}^{2n} \frac{\partial \sigma_{n,n}}{\partial u_k}\, \lambda^k
    &= \lambda^{2n} - 2\sum_{i=0}^{n-1} \alpha_i\, \lambda^{n+i}
       + \sum_{i,j=0}^{n-1} \alpha_i \alpha_j\, \lambda^{i+j} \nonumber \\
    &= \left(\lambda^n - \sum_{i=0}^{n-1} \alpha_i\, \lambda^i\right)^{\!2}
     = p_n(\lambda)^2.
    \label{eq:schur-gradient}
\end{align}
The Schur-complement gradient polynomial is the square of the monic orthogonal polynomial.
This identity has been verified numerically for $M = 5$ and $M = 7$ to a precision of $10^{-7}$.

For the closure~\eqref{eq:ri-general}, the Jacobian last-row entries are $C_{u_k} = \partial u_{2n}^{\min}/\partial u_k + \partial D/\partial u_k$ for $k = 0, \ldots, 2n-1$.
Using Eq.~\eqref{eq:schur-gradient} and the relation $\partial u_{2n}^{\min}/\partial u_k = -\partial \sigma_{n,n}/\partial u_k$ (for $k < 2n$), the characteristic polynomial~\eqref{eq:charpoly} becomes
\begin{equation}
    \boxed{P_D(\lambda) = p_n(\lambda)^2 - \mathcal{D}(\lambda),}
    \label{eq:structural-identity}
\end{equation}
where
\begin{equation}
    \mathcal{D}(\lambda) = \sum_{k=0}^{2n-1} \frac{\partial D}{\partial u_k}\, \lambda^k
    \label{eq:D-polynomial}
\end{equation}
is the derivative polynomial of the margin.
The Gramian boundary closure ($D = 0$) gives $P = p_n^2$, which has $n$ double roots.
Any positive margin splits these double roots by subtracting $\mathcal{D}$.
Whether the split produces real or complex roots depends on the polynomial structure of $\mathcal{D}$.

Any margin that depends on $(\rho, v, \theta)$ only is constrained by dimensional analysis and reflection symmetry.
Since $D$ has the same dimensions as $u_{2n}$ (the $2n$th velocity moment), it scales as $\rho\,\theta^n$ (degree one in density, degree $n$ in temperature).
The realizability gap $u_{2n} - u_{2n}^{\min} = \|p_n\|_f^2$ is invariant under $c \to -c$ (which sends $v \to -v$), so $D$ must be even in $v$.
The most general form is therefore $D = \rho\, \theta^n\, f(v^2/\theta)$ for a dimensionless function $f$.
Equilibrium preservation requires $f(0) = n!$.
At equilibrium $(v = 0)$, the chain rule through $\rho = u_0$, $v = u_1/u_0$, $\theta = u_2/u_0 - v^2$ gives
\begin{equation}
    \frac{\partial D}{\partial u_0}\bigg|_{\Eq} = -(n{-}1)\,n!\,,\quad
    \frac{\partial D}{\partial u_1}\bigg|_{\Eq} = 0,\quad
    \frac{\partial D}{\partial u_2}\bigg|_{\Eq} = n\cdot n!\,,
    \label{eq:covariance}
\end{equation}
regardless of $f$, because the $f'(0)$ terms carry a factor of $v$ that vanishes at $v = 0$.
The derivative polynomial~\eqref{eq:D-polynomial} at equilibrium is therefore
\begin{equation}
    \mathcal{D}(\lambda)\big|_{\Eq} = -(n{-}1)\,n! + n\cdot n!\,\lambda^2,
\end{equation}
a quadratic in $\lambda$ with no free parameters.
The equilibrium characteristic polynomial $P_D = p_n^2 - \mathcal{D}$ is thus uniquely determined for every equilibrium-preserving $(\rho,v,\theta)$-only margin, independent of the specific function $f$.

At the standard Gaussian equilibrium $(\rho, v, \theta) = (1, 0, 1)$ with $M = 5$ ($n = 3$), the constant-margin Jacobian has last row $(-12, 0, 9, 0, 6, 0)$, giving the characteristic polynomial
\begin{equation}
    P(\lambda) = \lambda^6 - 6\lambda^4 - 9\lambda^2 + 12.
    \label{eq:m5-failure}
\end{equation}
Substituting $\mu = \lambda^2$ yields $q(\mu) = \mu^3 - 6\mu^2 - 9\mu + 12$.
The product of the three $\mu$-roots is $-12$ by Vieta's formula.
Since the product is negative, at least one $\mu$ is negative, producing a conjugate pair of complex eigenvalues $\lambda = \pm i\sqrt{|\mu|}$.
The constant-margin closure is not hyperbolic at Gaussian equilibrium for $M = 5$.
Numerically, the three $\mu$-roots are approximately $7.04$, $0.89$, and $-1.92$: the negative root produces a conjugate imaginary pair $\lambda \approx \pm 1.387\,i$, confirming non-hyperbolicity at the most benign possible state.
The simplified Grad closure $C = g_6(\rho, v, \theta)$, which sets the closure equal to the full Gaussian sixth moment, has a different characteristic polynomial $\lambda^6 - 45\lambda^2 + 30$ and produces the larger imaginary pair $\lambda \approx \pm 2.649\,i$.
Both closures fail at equilibrium for $M = 5$, but through different characteristic polynomials.

For $M = 3$ ($n = 2$), the same calculation gives $\lambda^4 - 6\lambda^2 + 3$, with $\mu$-roots $3 \pm \sqrt{6}$.
Both are positive, so the constant-margin closure is hyperbolic at equilibrium for $M = 3$.
Table~\ref{tab:eq-jacobians} lists the constant-margin Jacobian last rows and their hyperbolicity status at standard equilibrium for $M = 3, 5, 7, 9$.
The equilibrium characteristic polynomial coefficients for the interpolated family are tabulated in Appendix~\ref{app:polynomials}.

\begin{table}
\caption{
Last row of the constant-margin RI Jacobian at the standard Gaussian equilibrium $(\rho, v, \theta) = (1, 0, 1)$ for $M = 3, 5, 7, 9$.
Only even entries are nonzero.
The column ``hyp.''\ indicates whether all eigenvalues are real.
}
\label{tab:eq-jacobians}
\begin{ruledtabular}
\begin{tabular}{cll}
$M$ & $\nabla C_{\RI}^0\big|_{\Eq}$ (even entries) & hyp. \\
\colrule
3  & $-3, \; 6$ & yes \\
5  & $-12, \; 9, \; 6$ & no \\
7  & $-81, \; 132, \; {-42}, \; 12$ & no \\
9  & $-480, \; 375, \; 300, \; {-130}, \; 20$ & no \\
\end{tabular}
\end{ruledtabular}
\end{table}

\section{Normalized Schur ratios and the geometric closure}
\label{sec:schur-ratios}

A margin that couples to higher moments requires features computed from $u_3, \ldots, u_{2n-1}$.
The natural features are the lower-order Schur complements, normalized by their Gaussian equilibrium values.

For $j \geq 1$, define the Schur complement at order $j$:
\begin{equation}
    \sigma_{j,j} = u_{2j} - u_{j,2j-1}^T H_{j-1}^{-1} u_{j,2j-1},
    \label{eq:sigma-j}
\end{equation}
where $u_{j,2j-1} = (u_j, \ldots, u_{2j-1})^T$ and $H_{j-1} = (u_{i+k})_{i,k=0}^{j-1}$.
In the realizability interior, $\sigma_{j,j} > 0$.
At Gaussian equilibrium, $\sigma_{j,j}^{\Eq} = \rho\, j!\, \theta^j$ by the Gaussian squared norm formula~\eqref{eq:eq-gap} applied at order $j$.
We define the normalized Schur ratio
\begin{equation}
    R_j(U) = \frac{\sigma_{j,j}(U)}{\rho\, j!\, \theta^j},
    \label{eq:R-j}
\end{equation}
and set $R_0 = R_1 = 1$ by convention.
At any Gaussian equilibrium, $R_j = 1$ for every $j$.
In the realizability interior, $R_j > 0$.

The ratio $R_j$ measures the distance from the realizability boundary at order $j$, relative to the Gaussian equilibrium value at the same density and temperature.
A value $R_j < 1$ means the moment state is closer to the realizability boundary than a Gaussian with the same $(\rho, v, \theta)$.
A value $R_j > 1$ means the state is farther into the interior than the Gaussian, which occurs for distributions with heavier tails or stronger bimodal separation.

To construct a margin that matches the Morin--McDonald equilibrium Jacobian, define the weights $w_k = 2k/[n(n-1)]$ for $k = 1, \ldots, n-1$, which sum to one, and the normalized Schur-complement ratios $q_k = R_k / R_{k-1}$.
If the margin $D$ is written as $D = \rho\, n!\, \theta^n \cdot \alpha(U)$ for some positive scalar function $\alpha$ with $\alpha(G) = 1$, the first variation of $\log \alpha$ at equilibrium must match that of the Morin--McDonald closure.
Since the $a$-dispersion term in the Morin--McDonald formula (the sum of squared recurrence-coefficient differences) has zero first variation at equilibrium, the matching condition involves only the $b_k$-part.
The weighted sum $\sum w_k\, \mathrm{d}\log q_k$ telescopes:
\begin{align}
    \sum_{k=1}^{n-1} w_k\, \mathrm{d}\log q_k
    &= \sum_{k=1}^{n-1} w_k\left(\mathrm{d}\log R_k - \mathrm{d}\log R_{k-1}\right) \nonumber \\
    &= \frac{2}{n}\, \mathrm{d}\log R_{n-1}
       - \frac{2}{n(n-1)} \sum_{j=2}^{n-2} \mathrm{d}\log R_j.
    \label{eq:telescoping}
\end{align}
The full first variation of $\log \alpha$ at equilibrium includes an additional $\mathrm{d}\log R_{n-1}$ from the outer prefactor $R_{n-1}$ in the Morin--McDonald formula.
Adding this term gives the unique power-law exponents
\begin{equation}
    p_{n-1} = \frac{n+2}{n}, \qquad
    p_j = -\frac{2}{n(n-1)} \quad (j = 2, \ldots, n-2).
    \label{eq:exponents}
\end{equation}
These have been verified by solving the overdetermined linear system $\sum_j p_j\, \nabla(\log R_j)\big|_G = (\nabla C_{\MM} - \nabla C_{\RI}^0)/\Delta_{2n}^{\Eq}\big|_G$ for $M = 5, 7, 9$, with residual norms below $10^{-9}$.

The sum of exponents is $1 + 4/[n(n-1)]$, which converges to 1 as $n \to \infty$.
At high moment orders, the closure simplifies: $\alpha \to R_{n-1}$, the ratio of the actual top Schur complement to its Gaussian value.

\section{The interpolated closure family}
\label{sec:family}

The closure combines three ingredients: a geometric mean of normalized Schur-complement ratios, an arithmetic mean, and an $a$-dispersion term that couples the top odd moment through the Chebyshev recurrence coefficients.

The Chebyshev forward algorithm applied to $(u_0, \ldots, u_{2n-1})$ produces recurrence coefficients $a_0, \ldots, a_{n-1}$ and $b_1, \ldots, b_{n-1}$, which determine the monic orthogonal polynomials via $p_{k+1}(c) = (c - a_k)p_k(c) - b_k p_{k-1}(c)$.
At Gaussian equilibrium, $a_k = v$ for all $k$ and $b_k = k\theta$.
The three ingredients are
\begin{align}
    G_n &= \prod_{k=1}^{n-1} q_k^{w_k}, \label{eq:G-n} \\
    A_n &= \sum_{k=1}^{n-1} w_k\, q_k, \label{eq:A-n} \\
    S_n &= \frac{1}{n(n-1)\theta} \sum_{\ell=0}^{n-1} (a_{n-1} - a_\ell)^2, \label{eq:S-n}
\end{align}
where $q_k = R_k / R_{k-1}$ and $w_k = 2k/[n(n-1)]$.
The interpolated closure is
\begin{equation}
    \boxed{
    C_{\eta,n}(U) = u_{2n}^{\min}(U) + \rho\, n!\, \theta^n\, R_{n-1}\!\left[(1-\eta)\, G_n + \eta\, A_n + S_n\right],
    }
    \label{eq:closure-family}
\end{equation}
with $\eta \in [0,1]$.
Explicit formulas for $M = 5$ and $M = 7$ are given in Appendix~\ref{app:explicit}.

At any Gaussian equilibrium, $R_j = 1$, $q_k = 1$, and $a_k = v$ for all $k$, so $G_n = A_n = 1$ and $S_n = 0$.
The margin reduces to $\rho\, n!\, \theta^n$, which is the Gaussian squared norm~\eqref{eq:eq-gap}.
The closure reproduces the exact Gaussian moment for every $\eta$.

The first variations of $G_n$ and $A_n$ coincide at equilibrium because $G_n = A_n = 1$ and $\mathrm{d}G_n = \sum w_k\, \mathrm{d}q_k = \mathrm{d}A_n$ when all $q_k = 1$.
The $S_n$ term has zero first variation at equilibrium because it is quadratic in the differences $a_{n-1} - a_\ell$, all of which vanish there.
Every member of the family therefore has the same equilibrium Jacobian, which matches the Morin--McDonald Jacobian.
This has been verified numerically for $M = 5, 7, 9$ with maximum discrepancy below $1.2 \times 10^{-7}$.

The equilibrium Jacobian matching extends beyond the interpolated family.
Let $\Delta = \rho\, n!\, \theta^n$ and write $C = u_{2n}^{\min} + \Delta\, \gamma(U)$ with $\gamma(G) = 1$.
At a Gaussian equilibrium $G$, the product rule gives
\begin{equation}
    \nabla C\big|_G = \nabla u_{2n}^{\min}\big|_G + \nabla\Delta\big|_G + \Delta(G)\, \nabla\gamma\big|_G.
    \label{eq:universality}
\end{equation}
The first two terms are common to every such closure.
Hence two closures with $\gamma(G) = 1$ have the same equilibrium Jacobian if and only if they have the same covector $\nabla\gamma\big|_G$.
The constant-margin closure ($\gamma \equiv 1$) has $\nabla\gamma\big|_G = 0$ and produces the non-hyperbolic Jacobian of Sec.~\ref{sec:structural}.
The interpolated family has a common nonzero covector $\nabla\gamma_\eta\big|_G = \nabla\gamma_{\MM}\big|_G$, because $\mathrm{d}G_n = \mathrm{d}A_n$ and $\mathrm{d}S_n = 0$ at equilibrium (shown above), and this shared value equals the Morin--McDonald gradient by the exponent matching of Sec.~\ref{sec:schur-ratios}.
All members of the family therefore share the Morin--McDonald equilibrium Jacobian, not the constant-margin one.
The Grad closure~\cite{Grad1949} has a different $\nabla\gamma\big|_G$ and produces a different equilibrium Jacobian for $M \geq 5$.

At $\eta = 1$, the margin becomes $\rho\, n!\, \theta^n\, R_{n-1}(A_n + S_n)$.
Since $\rho\, n!\, \theta^n R_{n-1} = n\theta\, \sigma_{n-1,n-1}$, the total margin at $\eta = 1$ is $n\theta\,\sigma_{n-1,n-1}(A_n + S_n)$.
Converting $A_n$ and $S_n$ back to recurrence coefficients using $n\theta\, A_n = (2/(n-1))\sum b_k$ and $n\theta\, S_n = (1/(n-1))\sum(a_{n-1}-a_\ell)^2$, the total margin becomes $\beta_n\, \sigma_{n-1,n-1}$ with
\begin{equation}
    \beta_n = \frac{2}{n-1}\sum_{k=1}^{n-1} b_k + \frac{1}{n-1}\sum_{\ell=0}^{n-1}(a_{n-1}-a_\ell)^2.
    \label{eq:beta-n}
\end{equation}
This is the Morin--McDonald closure.
In the notation of~\cite{MorinMcDonald2025}, the even-order B-hierarchy closes with the recurrence coefficient $b_{(n+1)/2} = \frac{4}{n-1}\sum b_k + \frac{2}{n-1}\sum(a_{(n-1)/2} - a_l)^2$ (their Eq.~(48), with $n$ denoting the number of moments).
The odd-order formula~\eqref{eq:beta-n} has the same structure with coefficients halved: $4/(n{-}1) \to 2/(n{-}1)$ and $2/(n{-}1) \to 1/(n{-}1)$, reflecting the different role of $\beta_n$ (it multiplies $\sigma_{n-1,n-1}$ instead of $\sigma_{n-2,n-2}$ in the even case).
The identity $C_{1,n} = C_{\MM,n}$ has been verified to machine precision across the full bimodal benchmark for $M = 5, 7, 9$.

The closure is affine in $\eta$:
\begin{equation}
    C_{\eta,n} = (1-\eta)\, C_{0,n} + \eta\, C_{1,n}.
    \label{eq:affine}
\end{equation}
By the weighted arithmetic-geometric mean inequality, $A_n \geq G_n$ with equality only when all $q_k$ are equal.
The margin $D_\eta = C_\eta - u_{2n}^{\min}$ is monotonically increasing in $\eta$:
\begin{equation}
    D_{0,n}(U) \leq D_{\eta,n}(U) \leq D_{1,n}(U).
    \label{eq:amgm-order}
\end{equation}
The geometric endpoint ($\eta = 0$) provides the smallest margin.
The arithmetic endpoint ($\eta = 1$, Morin--McDonald) provides the largest.
This ordering has been verified with zero violations across 71 bimodal benchmark states for $M = 5, 7, 9$.

The characteristic polynomial is also affine in $\eta$: $P_\eta = (1-\eta)P_0 + \eta P_1$.
On the bimodal benchmark, both $P_0$ and $P_1$ are real-rooted at every tested state, and all intermediate $\eta$ values are also real-rooted across all 71 states for $M = 5, 7, 9$.
The roots of $P_0$ and $P_1$ are close, since the endpoints differ at second order in the spread of normalized Schur-complement ratios, but they do not strictly interlace, so the all-$\eta$ result is a numerical observation.

For $M = 3$ ($n = 2$), only $R_0$ and $R_1$ are available, both equal to 1.
The geometric and arithmetic means are both 1, so the $\eta$-interpolation makes no difference.
Only the $a$-dispersion $S_2 = (a_1 - a_0)^2/(2\theta)$ contributes, and the family reduces to the Morin--McDonald closure for all $\eta$.
The constant-margin closure (without $S_2$) is a different closure and is slightly more accurate on the bimodal benchmark at $M = 3$ ($\max|e| = 0.135$ versus $0.144$), because the $a$-dispersion over-corrects at this low order.

\section{Entropy structure}
\label{sec:entropy}

A smooth entropy pair $(h, \psi)$ for the closed moment system~\eqref{eq:closed-system} satisfies $\partial_t h + \partial_x \psi \leq 0$ with $h$ strictly convex.
The Godunov--Mock compatibility condition requires $\nabla^2 h \cdot A$ to be symmetric, where $A$ is the flux Jacobian~\eqref{eq:companion-jacobian}.
We analyze the entropy structure at three levels.

\subsection{Linearized entropy}
\label{sec:linearized-entropy}

At any Maxwellian equilibrium, the Jacobian $A^0$ has $N = M+1$ distinct real eigenvalues $\lambda_1 < \cdots < \lambda_N$ for every tested $M$ (Table~\ref{tab:poly}, verified through $M = 17$).
The Vandermonde matrix $V_{ki} = \lambda_i^k$ diagonalizes $A^0$, and the family of matrices
\begin{equation}
    S^0 = V^{-T} D\, V^{-1}, \qquad D = \diag(d_1, \ldots, d_N), \quad d_i > 0,
    \label{eq:symmetrizer}
\end{equation}
provides a symmetrizer for each choice of positive weights $D$.
Since $S^0$ is congruent to the positive definite matrix $D$, it is itself positive definite.
The product $S^0 A^0 = V^{-T} D \Lambda\, V^{-1}$ is symmetric because $D\Lambda$ is diagonal.
The quadratic entropy $h_2(U) = \frac{1}{2}(U - U^{\Eq})^T S^0 (U - U^{\Eq})$ therefore yields a valid entropy inequality near any Maxwellian equilibrium.
Because the equilibrium Jacobian is independent of $\eta$ (Sec.~\ref{sec:family}), this linearized entropy exists for every member of the interpolated family.

\subsection{BGK dissipation near equilibrium}
\label{sec:bgk-dissipation}

Let $\mathcal{G}(U) = (g_0, g_1, \ldots, g_M)^T$ denote the Maxwellian moment projection and let
\begin{equation}
    P = \frac{\partial \mathcal{G}}{\partial U}\bigg|_G, \qquad Q = I - P,
    \label{eq:maxwellian-proj}
\end{equation}
where $G = U^{\Eq}$.
Since $\mathcal{G} \circ \mathcal{G} = \mathcal{G}$, $P$ is idempotent ($P^2 = P$), and $\operatorname{Range}(P)$ is the tangent space to the Maxwellian manifold.
The BGK source satisfies
\begin{equation}
    S(U) = \frac{\mathcal{G}(U) - U}{\tau} = -\frac{1}{\tau}\, Q\, \delta U + O(|\delta U|^2), \qquad \delta U = U - G.
    \label{eq:bgk-linearized}
\end{equation}
The linearized entropy production is therefore
\begin{equation}
    \nabla h_2 \cdot S = -\frac{1}{\tau}\, \delta U^T S^0\, Q\, \delta U + O(|\delta U|^3) = -\frac{1}{2\tau}\, \delta U^T \!\left(S^0 Q + Q^T S^0\right) \delta U + O(|\delta U|^3).
    \label{eq:bgk-dissipation}
\end{equation}
This leading quadratic form is nonpositive for all perturbations if and only if
\begin{equation}
    P^T S^0\, Q = 0, \qquad \text{equivalently} \quad S^0 P = P^T S^0.
    \label{eq:source-compat}
\end{equation}
Under this source-compatibility condition, $\delta U^T S^0\, Q\, \delta U = \|Q\, \delta U\|_{S^0}^2$ and
\begin{equation}
    \nabla h_2 \cdot S = -\frac{1}{\tau}\, \|Q\, \delta U\|_{S^0}^2 + O(|\delta U|^3),
    \label{eq:bgk-compatible}
\end{equation}
which is negative semidefinite with kernel equal to the three-dimensional Maxwellian tangent space $\operatorname{Range}(P)$.
BGK dissipation thus requires a source-compatible choice of the diagonal weights $D$ and does not follow from $S^0 > 0$ alone.
Source-compatible weights have been found by linear programming for $M = 3, 5, 7, 9$.

\subsection{Nonlinear entropy}
\label{sec:nonlinear-entropy}

A smooth (nonlinear) entropy $h(U)$ in a neighborhood of equilibrium requires the Hessian field $H(U) = \nabla^2 h(U)$ to satisfy the Godunov--Mock compatibility condition at every $U$, not only at equilibrium.
Expanding $H = S^0 + T_k\, \delta U_k + O(|\delta U|^2)$, the first-order correction $T_k$ must satisfy the Sylvester equation
\begin{equation}
    T_k\, A^0 - (A^0)^T T_k = \varphi\, C_k^T - C_k\, \varphi^T,
    \label{eq:sylvester}
\end{equation}
for each perturbation direction $k = 0, \ldots, M$, where $\varphi = S^0 e_N$ is the last column of $S^0$ and $C_{ik} = \partial^2 C / \partial u_i\, \partial u_k$ is the closure Hessian.
The right-hand side is antisymmetric by construction.
In the eigenbasis of $A^0$, the off-diagonal elements of $T_k$ are uniquely determined, while the $N$ diagonal elements per direction are free parameters, giving $N^2$ free unknowns in addition to the $N$ weights $d_1, \ldots, d_N$.

A smooth entropy exists in a neighborhood of equilibrium if and only if the Poincar\'{e} integrability condition
\begin{equation}
    T_{ijk} = T_{ikj}
    \label{eq:integrability}
\end{equation}
is satisfiable: the resulting homogeneous linear system in $(d_1, \ldots, d_N, T_{11,1}, \ldots, T_{NN,N})$ must admit a solution with all $d_i > 0$.

Table~\ref{tab:entropy} summarizes the computation.
For $M = 3$, the null space has dimension~6 among 20 unknowns, and the $D$-projection intersects the positive orthant (verified by linear programming).
A smooth entropy therefore exists for $M = 3$, where the closure is independent of $\eta$.
For $M = 5$ and $M = 7$, the linear programming feasibility check certifies that no $D > 0$ solution exists.
The obstruction is independent of $\eta$ (tested at $\eta = 0, 0.5, 1$) and independent of the equilibrium state (tested at four distinct $(\rho, \theta)$ combinations).

\begin{table}
\caption{Integrability of the nonlinear entropy extension.
The system~\eqref{eq:integrability} produces a homogeneous linear system in the symmetrizer weights $D$ and diagonal perturbation parameters $T_{\mathrm{diag}}$.
The column ``$D > 0$?''\ indicates whether any solution has all weights positive, determined by linear programming.
The result is independent of $\eta$.}
\label{tab:entropy}
\begin{ruledtabular}
\begin{tabular}{ccccc}
$M$ & equations & unknowns & null dim & $D > 0$? \\
\colrule
3 & 24 & 20 & 6 & yes \\
5 & 90 & 42 & 6 & no \\
7 & 224 & 72 & 8 & no \\
\end{tabular}
\end{ruledtabular}
\end{table}

The obstruction at $M = 5$ and $M = 7$ is structural: it arises from the companion-type Jacobian and the closure Hessian, not from a specific choice of $\eta$ or equilibrium state.
No smooth entropy is known for any odd-$M$ closure with $M \geq 5$, including the Morin--McDonald closure (which has the same obstruction, since it is a member of the interpolated family at $\eta = 1$).
The maximum-entropy closure for odd $M$ is likewise degenerate: by the theorem of Junk~\cite{Junk1998}, the maximum-entropy distribution on the real line degenerates to an $n$-point discrete measure, placing the extended moment vector on the realizability boundary.

\section{Numerical verification}
\label{sec:numerics}

All Jacobians in this section are computed by complex-step differentiation (Appendix~\ref{app:complex-step}): $\partial C/\partial u_j = \mathrm{Im}[C(u + ih\, e_j)]/h$ with $h = 10^{-30}$.
This method has no cancellation error and gives machine-precision derivatives regardless of Hankel conditioning.

\subsection{Bimodal benchmark}
\label{sec:bimodal}

\begin{figure}
\centering
\includegraphics[width=0.9\columnwidth]{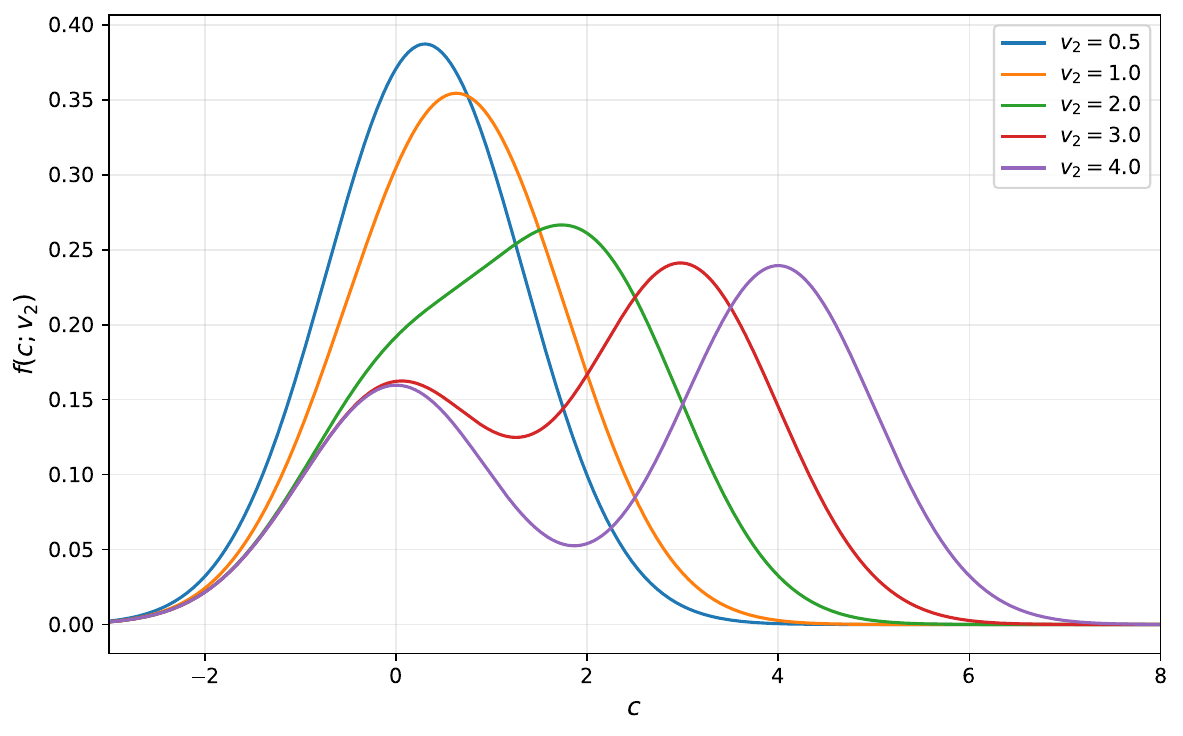}
\caption{The bimodal benchmark distribution~\eqref{eq:bimodal} for five values of the velocity separation $v_2$. At $v_2 = 0.5$ the distribution is nearly Gaussian. As $v_2$ increases, two distinct peaks form with the right peak carrying 60\% of the mass.}
\label{fig:bimodal-distributions}
\end{figure}

The benchmark distribution of Yilmaz \textit{et al.}~\cite{YilmazRGD2024} is a superposition of two Gaussians:
\begin{equation}
    f(c; v_2) = 0.4\,\mathcal{M}(c; 0, 1) + 0.6\,\mathcal{M}(c; v_2, 1),
    \label{eq:bimodal}
\end{equation}
with $v_2 \in [0.5, 4]$ in steps of 0.05 (71 states).
Figure~\ref{fig:bimodal-distributions} shows the distribution for several values of $v_2$.
Reference moments are computed analytically from the Gaussian recurrence.
Table~\ref{tab:bimodal} gives the maximum and mean absolute relative errors and the number of complex eigenvalues for $\eta = 0$, $0.5$, $1$ and $M = 5, 7, 9$.
Table~\ref{tab:comparison} compares the interpolated family against the Gramian (boundary) closure, the Grad Hermite expansion closure~\cite{Grad1949}, and the Morin--McDonald closure for $M = 5, 7, 9, 11$.

\begin{figure}
\centering
\includegraphics[width=0.9\columnwidth]{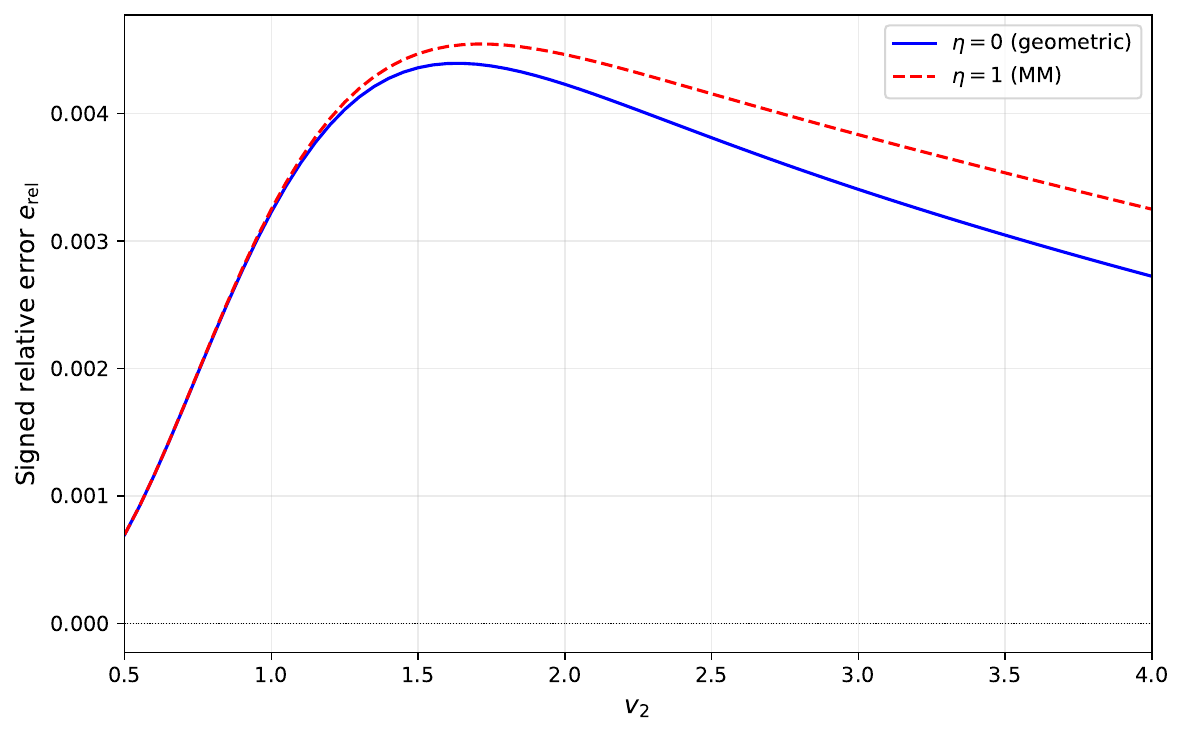}
\caption{Signed relative error of the predicted eighth moment versus $v_2$ for $M = 7$. The geometric ($\eta = 0$, solid) and arithmetic ($\eta = 1$, dashed) endpoints of the interpolated family are both hyperbolic at every state. The geometric endpoint produces a consistently smaller error for $v_2 > 1.2$, with a maximum advantage of about 5\% in the mid-range.}
\label{fig:error-comparison}
\end{figure}

All members of the family have zero hyperbolicity failures on this benchmark for $M = 5, 7, 9$.
The geometric endpoint ($\eta = 0$) is 2--4\% more accurate than the arithmetic (Morin--McDonald) endpoint ($\eta = 1$), consistent with the AM--GM ordering~\eqref{eq:amgm-order}: the minimal margin tracks the reference more closely.
Figure~\ref{fig:error-comparison} compares the signed relative error across the benchmark for $M = 7$.

\begin{table}
\caption{
Bimodal benchmark accuracy and hyperbolicity for the interpolated closure at three values of $\eta$ and three moment orders.
The column $N_\mathbb{C}$ counts states with at least one complex eigenvalue out of 71 total.
}
\label{tab:bimodal}
\begin{ruledtabular}
\begin{tabular}{ccrrc}
$M$ & $\eta$ & $\max|e_{\mathrm{rel}}|$ & $\mathrm{mean}|e_{\mathrm{rel}}|$ & $N_\mathbb{C}$ \\
\colrule
5 & 0   & 0.0127 & 0.0079 & 0/71 \\
5 & 0.5 & 0.0129 & 0.0085 & 0/71 \\
5 & 1 (MM) & 0.0132 & 0.0091 & 0/71 \\
\colrule
7 & 0   & 0.0044 & 0.0034 & 0/71 \\
7 & 0.5 & 0.0045 & 0.0035 & 0/71 \\
7 & 1 (MM) & 0.0045 & 0.0037 & 0/71 \\
\colrule
9 & 0   & 0.0017 & 0.0010 & 0/71 \\
9 & 0.5 & 0.0017 & 0.0010 & 0/71 \\
9 & 1 (MM) & 0.0017 & 0.0011 & 0/71 \\
\end{tabular}
\end{ruledtabular}
\end{table}

\begin{figure}
\centering
\includegraphics[width=\columnwidth]{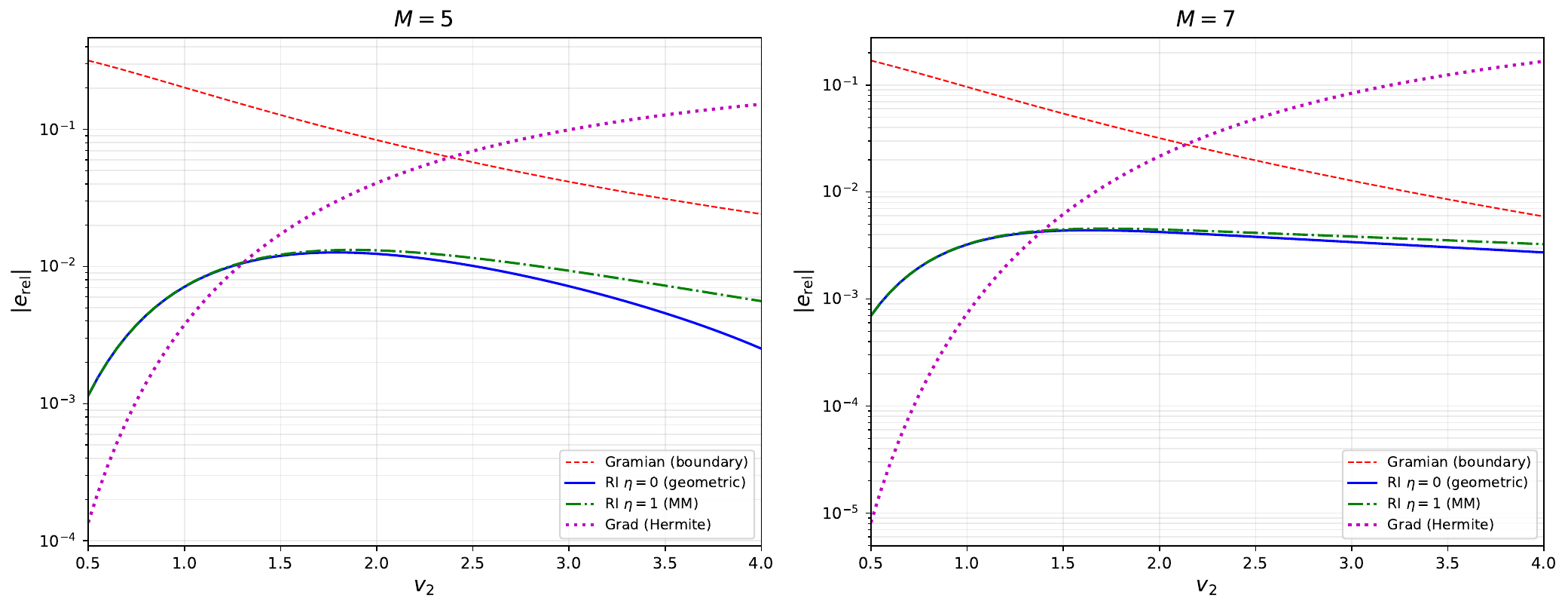}
\caption{Absolute relative closure error on the bimodal benchmark for $M = 5$ (left) and $M = 7$ (right), comparing the Gramian boundary closure (dashed red), the Grad Hermite expansion closure (dotted magenta), and the interpolated family endpoints $\eta = 0$ (solid blue) and $\eta = 1$/MM (dash-dotted green). The interpolated family errors are 10--40$\times$ smaller than either the Gramian or Grad closure across the benchmark.}
\label{fig:closure-comparison}
\end{figure}

The Gramian closure, which saturates the realizability boundary, has errors 10--30$\times$ larger than the interpolated family and does not converge with $M$ at the same rate.
Its eigenvalues are real (the $n$ roots of the monic orthogonal polynomial, each doubled) but the Jacobian is not diagonalizable.
The Grad Hermite expansion closure, which projects the distribution onto a polynomial perturbation of the local Maxwellian, produces errors of 15--20\% on the bimodal benchmark for all $M$ tested.
Strikingly, the Grad error does not decrease with moment order: at $M = 11$, the maximum relative error is $0.20$, worse than at $M = 5$ ($0.15$).
This divergent behavior reflects the finite convergence radius of the Hermite expansion for strongly bimodal distributions.
Furthermore, the Grad closure loses strict hyperbolicity on 31--45\% of bimodal states (Table~\ref{tab:comparison}), while the interpolated family has zero hyperbolicity failures on the same benchmark.
Figure~\ref{fig:closure-comparison} compares the absolute closure error across all four closures for $M = 5$ and $M = 7$.

\begin{table}
\caption{
Bimodal benchmark: closure accuracy and hyperbolicity comparison for four closures at $M = 5, 7, 9, 11$.
The Gramian has weakly hyperbolic eigenvalues (real but not diagonalizable).
The Grad (Hermite expansion) closure loses strict hyperbolicity on 31--45\% of states.
The interpolated family ($\eta = 0$ and $\eta = 1$) has zero hyperbolicity failures.
}
\label{tab:comparison}
\begin{ruledtabular}
\begin{tabular}{clrrc}
$M$ & Closure & $\max|e_{\mathrm{rel}}|$ & $\mathrm{mean}|e_{\mathrm{rel}}|$ & $N_\mathbb{C}$ \\
\colrule
5 & Gramian          & 0.3174 & 0.1015 & 0/71$^\dag$ \\
5 & RI $\eta{=}0$    & 0.0127 & 0.0079 & 0/71 \\
5 & RI $\eta{=}1$ (MM) & 0.0132 & 0.0091 & 0/71 \\
5 & Grad             & 0.1525 & 0.0619 & 22/71 \\
\colrule
7 & Gramian          & 0.1692 & 0.0440 & 0/71$^\dag$ \\
7 & RI $\eta{=}0$    & 0.0044 & 0.0034 & 0/71 \\
7 & RI $\eta{=}1$ (MM) & 0.0045 & 0.0037 & 0/71 \\
7 & Grad             & 0.1664 & 0.0525 & 32/71 \\
\colrule
9 & Gramian          & 0.0880 & 0.0196 & 0/71$^\dag$ \\
9 & RI $\eta{=}0$    & 0.0017 & 0.0010 & 0/71 \\
9 & RI $\eta{=}1$ (MM) & 0.0017 & 0.0011 & 0/71 \\
9 & Grad             & 0.1817 & 0.0461 & 29/71 \\
\colrule
11 & Gramian         & 0.0450 & 0.0089 & 0/71$^\dag$ \\
11 & RI $\eta{=}0$   & 0.0007 & 0.0004 & 0/71 \\
11 & RI $\eta{=}1$ (MM) & 0.0007 & 0.0004 & 0/71 \\
11 & Grad            & 0.2002 & 0.0421 & 23/71 \\
\end{tabular}
\end{ruledtabular}
\begin{flushleft}
{\small $^\dag$\,Gramian eigenvalues are real but repeated ($n$ double roots); the Jacobian is not diagonalizable.}
\end{flushleft}
\end{table}

\subsection{Convergence with moment order}
\label{sec:convergence}

\begin{figure}
\centering
\includegraphics[width=0.85\columnwidth]{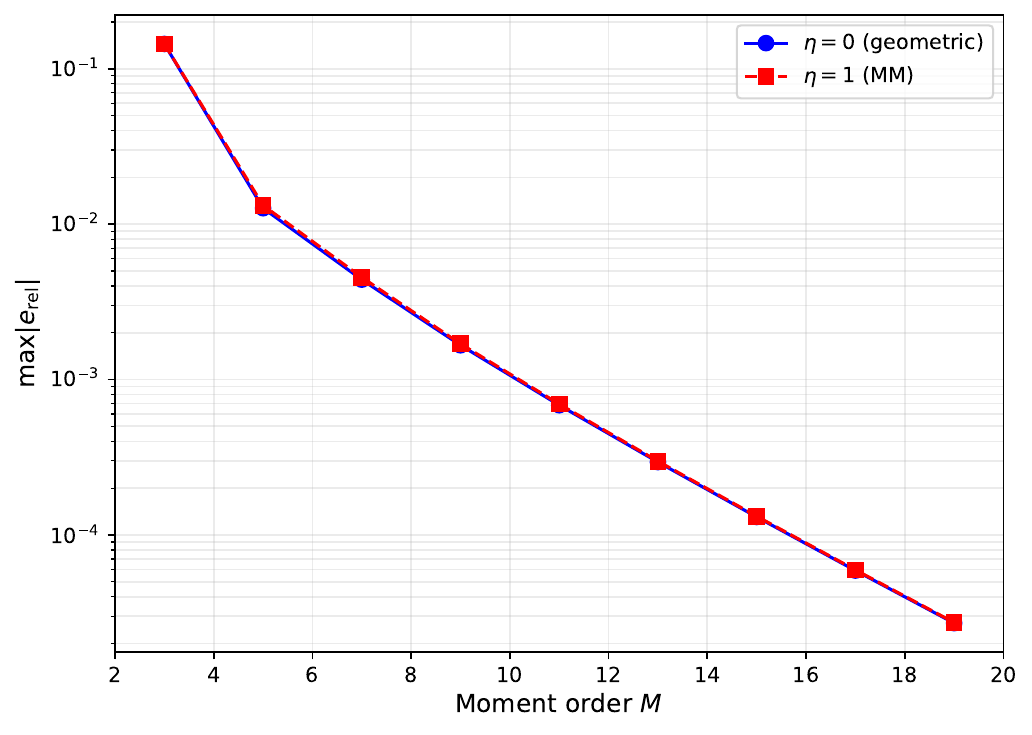}
\caption{Maximum absolute relative error on the bimodal benchmark versus moment order $M$ for the geometric ($\eta = 0$, circles) and arithmetic ($\eta = 1$, squares) endpoints. The two curves converge as $M$ increases, reaching a ratio of 0.99 at $M = 19$.}
\label{fig:convergence}
\end{figure}

Table~\ref{tab:convergence} lists the maximum absolute relative error on the bimodal benchmark for $M = 3, 5, \ldots, 19$ at $\eta = 0$ and $\eta = 1$.
The log-log slope of the error decay increases from 4.8 (between $M = 3$ and $M = 5$) to 7.0 (between $M = 17$ and $M = 19$), indicating convergence faster than any single power law $M^{-p}$.
The error ratio $e(\eta{=}0)/e(\eta{=}1)$ converges to 1 from below, reaching 0.99 at $M = 19$.
At high orders, the geometric and arithmetic endpoints become practically indistinguishable.
All computations use IEEE double precision.
The inner Hankel condition number reaches $\sim 10^{14}$ at $M = 19$, but the complex-step Jacobian is unaffected by Hankel conditioning, and the moment errors remain well-resolved.
Figure~\ref{fig:convergence} shows the error decay on a log-log scale.

\begin{table}
\caption{
Maximum absolute relative error on the bimodal benchmark versus moment order $M$, for the geometric ($\eta = 0$) and arithmetic ($\eta = 1$, Morin--McDonald) endpoints.
}
\label{tab:convergence}
\begin{ruledtabular}
\begin{tabular}{crrc}
$M$ & $\max|e|$ ($\eta{=}0$) & $\max|e|$ ($\eta{=}1$) & ratio \\
\colrule
 3 & $1.44 \times 10^{-1}$ & $1.44 \times 10^{-1}$ & 1.000 \\
 5 & $1.27 \times 10^{-2}$ & $1.32 \times 10^{-2}$ & 0.959 \\
 7 & $4.39 \times 10^{-3}$ & $4.55 \times 10^{-3}$ & 0.966 \\
 9 & $1.66 \times 10^{-3}$ & $1.70 \times 10^{-3}$ & 0.977 \\
11 & $6.84 \times 10^{-4}$ & $6.95 \times 10^{-4}$ & 0.984 \\
13 & $2.94 \times 10^{-4}$ & $2.98 \times 10^{-4}$ & 0.987 \\
15 & $1.30 \times 10^{-4}$ & $1.32 \times 10^{-4}$ & 0.989 \\
17 & $5.89 \times 10^{-5}$ & $5.94 \times 10^{-5}$ & 0.992 \\
19 & $2.71 \times 10^{-5}$ & $2.73 \times 10^{-5}$ & 0.991 \\
\end{tabular}
\end{ruledtabular}
\end{table}

\subsection{Random Gaussian-mixture stress test}
\label{sec:random}

We generate 1000 random realizable moment vectors for each $M$ by constructing Gaussian mixtures with 1--4 components, random weights (Dirichlet-distributed, scaled by a uniform factor in $[0.2, 3]$), random velocities ($v \sim \mathcal{N}(0,2)$), and random temperatures ($\log\theta \sim \mathcal{N}(0, 0.8)$).
Moments are computed analytically from the Gaussian recurrence.
The random seed is 42.

Table~\ref{tab:random} gives the hyperbolicity failure counts for $\eta = 0$, $0.5$, and $1$.
At $M = 5$, the geometric endpoint ($\eta = 0$) has 10 failures out of 1000 (1.0\%), while the Morin--McDonald endpoint ($\eta = 1$) has none.
At $M = 7$, $\eta = 0$ has 2 failures and $\eta = 1$ has none.
At $M = 9$, all closures share one failure at a near-boundary state with Hankel condition number $3.2 \times 10^{12}$, and $\eta = 0$ has one additional failure at a moderately conditioned state.

The $M = 5$ failure states cluster at extreme values of $R_2$: five states have $R_2 < 0.26$ (near the realizability boundary) and five have $R_2 > 2.8$ (deep in the non-Gaussian interior).
The success population has mean $R_2 \approx 1.0$.
The geometric mean produces too small a margin when $R_2 \ll 1$ and too large a margin when $R_2 \gg 1$.
The arithmetic mean provides more uniform coverage because it is less sensitive to individual $q_k$ outliers.

\begin{table}
\caption{
Random Gaussian-mixture stress test: number of states with at least one complex eigenvalue (strict hyperbolicity failure), out of 1000 random realizable states per $M$.
Jacobians computed by complex-step differentiation with threshold $|\mathrm{Im}\,\lambda| > 10^{-10}$.
Coincident real eigenvalues (weak hyperbolicity) were not observed at any tested state.
}
\label{tab:random}
\begin{ruledtabular}
\begin{tabular}{cccc}
$M$ & $\eta = 0$ & $\eta = 0.5$ & $\eta = 1$ (MM) \\
\colrule
5 & 10 (1.0\%) & 2 (0.2\%) & 0 (0.0\%) \\
7 & 2 (0.2\%) & 0 (0.0\%) & 0 (0.0\%) \\
9 & 2 (0.2\%) & 1 (0.1\%) & 1 (0.1\%) \\
\end{tabular}
\end{ruledtabular}
\end{table}

\subsection{The $R_j$ diagnostic}
\label{sec:diagnostic}

\begin{figure}
\centering
\includegraphics[width=0.9\columnwidth]{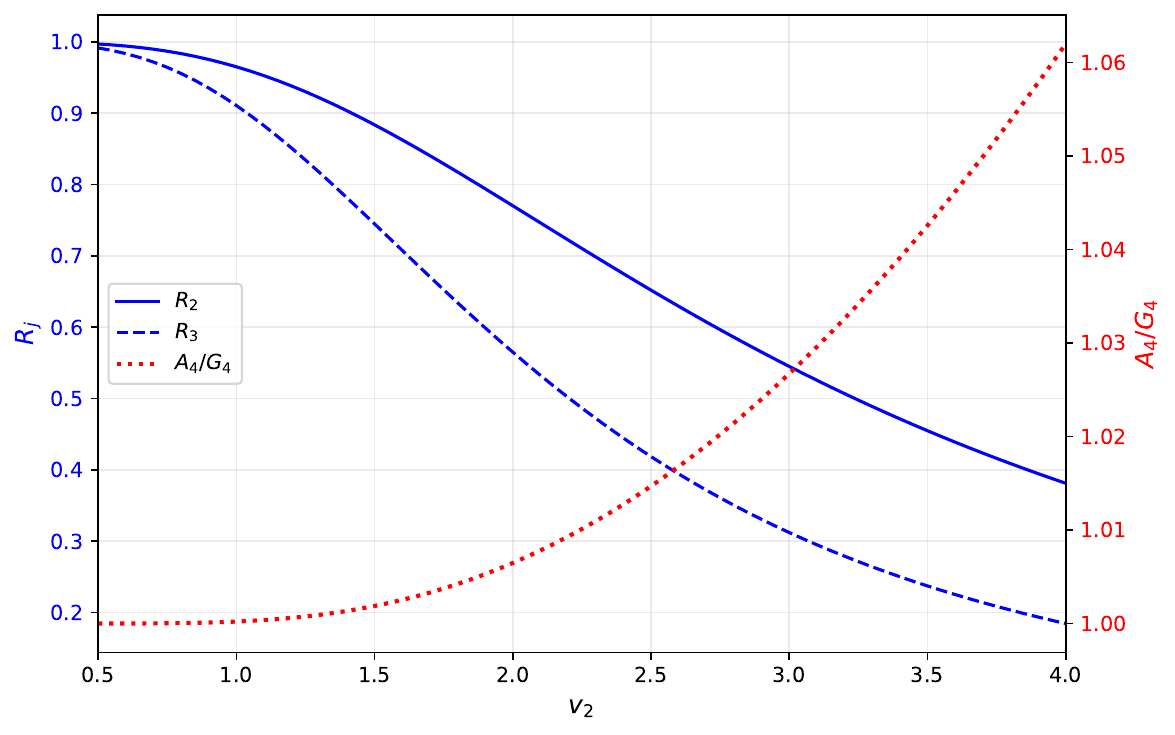}
\caption{Normalized Schur ratios $R_2$ (solid) and $R_3$ (dashed) and the AM--GM ratio $A_4/G_4$ (dotted, right axis) on the bimodal benchmark for $M = 7$. As the bimodal separation $v_2$ increases, $R_j$ decreases (the state moves toward the boundary) and the AM--GM ratio grows from 1.00 to 1.06.}
\label{fig:rj-variation}
\end{figure}

Table~\ref{tab:rj} shows how the normalized Schur ratios $R_2$ and $R_3$, the geometric and arithmetic means $G_4$ and $A_4$, and their ratio vary across the bimodal benchmark for $M = 7$.
Near equilibrium ($v_2 = 0.5$), $R_j \approx 1$ and $A/G \approx 1$.
As the bimodal separation $v_2$ increases, $R_j$ decreases (the state moves closer to the boundary), and the AM--GM ratio $A_4/G_4$ grows from 1.00 to 1.06.
This 6\% gap at the Schur-ratio level translates to the 3--5\% accuracy difference between the endpoints in Table~\ref{tab:bimodal}.
Figure~\ref{fig:rj-variation} shows the variation of $R_2$, $R_3$, and $A_4/G_4$ across the benchmark.

\begin{table}
\caption{
Normalized Schur ratios and their means on the bimodal benchmark for $M = 7$ ($n = 4$) at selected values of $v_2$.
}
\label{tab:rj}
\begin{ruledtabular}
\begin{tabular}{crrccc}
$v_2$ & $R_2$ & $R_3$ & $G_4$ & $A_4$ & $A_4/G_4$ \\
\colrule
0.5 & 0.997 & 0.991 & 0.996 & 0.996 & 1.000 \\
1.0 & 0.965 & 0.911 & 0.960 & 0.960 & 1.000 \\
1.5 & 0.884 & 0.745 & 0.881 & 0.883 & 1.002 \\
2.0 & 0.770 & 0.565 & 0.785 & 0.790 & 1.006 \\
2.5 & 0.652 & 0.419 & 0.695 & 0.705 & 1.015 \\
3.0 & 0.545 & 0.312 & 0.618 & 0.635 & 1.027 \\
3.5 & 0.455 & 0.237 & 0.556 & 0.579 & 1.043 \\
4.0 & 0.381 & 0.185 & 0.505 & 0.536 & 1.062 \\
\end{tabular}
\end{ruledtabular}
\end{table}

\subsection{Mott--Smith shock benchmark}
\label{sec:mott-smith}

The Mott--Smith distribution~\cite{MottSmith1951} for a normal shock at Mach number $\mathrm{Ma}$ with adiabatic exponent $\gamma$ is a spatial mixture of upstream and downstream Maxwellians:
\begin{equation}
    f(c; \phi) = (1-\phi)\,\mathcal{M}(c; v_1, \theta_1) + \phi\,\mathcal{M}(c; v_2, \theta_2),
    \label{eq:mott-smith}
\end{equation}
where the upstream state is $(\rho_1, v_1, \theta_1) = (1, \mathrm{Ma}\sqrt{\gamma}, 1)$ and the downstream state $(\rho_2, v_2, \theta_2)$ is determined by the Rankine--Hugoniot conditions.
The mixing parameter $\phi \in [0,1]$ represents position through the shock.
We use $\mathrm{Ma} = 4$ and $\gamma = 5/3$, following the test case of Yilmaz \textit{et al.}~\cite{YilmazSIAM2026}, giving $\rho_2 = 3.368$, $v_2 = 1.533$, $\theta_2 = 5.863$.

Table~\ref{tab:mott-smith} gives the maximum and mean relative closure errors across 201 uniformly spaced values of $\phi$.
All states are realizable for both $M = 5$ and $M = 7$.
At $M = 5$, the geometric endpoint is 8\% more accurate than the Morin--McDonald endpoint in max error.
At $M = 7$, the difference is negligible.
The margin ratio $D_1/D_0$ ranges from 1.000 to 1.077 ($M = 5$) and 1.000 to 1.052 ($M = 7$), confirming the AM--GM ordering.

\begin{table}
\caption{Mott--Smith shock closure accuracy ($\mathrm{Ma} = 4$, $\gamma = 5/3$, 201 mixing parameters $\phi$).}
\label{tab:mott-smith}
\begin{ruledtabular}
\begin{tabular}{ccrr}
$M$ & $\eta$ & $\max|e_{\mathrm{rel}}|$ & $\mathrm{mean}|e_{\mathrm{rel}}|$ \\
\colrule
5 & 0   & $1.23 \times 10^{-2}$ & $8.57 \times 10^{-3}$ \\
5 & 0.5 & $1.29 \times 10^{-2}$ & $8.93 \times 10^{-3}$ \\
5 & 1 (MM) & $1.35 \times 10^{-2}$ & $9.29 \times 10^{-3}$ \\
\colrule
7 & 0   & $3.87 \times 10^{-3}$ & $1.73 \times 10^{-3}$ \\
7 & 0.5 & $3.88 \times 10^{-3}$ & $1.72 \times 10^{-3}$ \\
7 & 1 (MM) & $3.88 \times 10^{-3}$ & $1.72 \times 10^{-3}$ \\
\end{tabular}
\end{ruledtabular}
\end{table}

On the Mott--Smith benchmark, the Gramian closure has maximum errors of $0.169$ ($M = 5$) and $0.078$ ($M = 7$).
The Grad closure has maximum errors of $0.086$ ($M = 5$) and $0.097$ ($M = 7$), again failing to converge with moment order.
Both endpoints of the interpolated family have maximum errors of $0.012$--$0.014$ ($M = 5$) and $0.004$ ($M = 7$), an order of magnitude better than either Gramian or Grad on this second independent benchmark.

\subsection{Equilibrium spectral radii}
\label{sec:spectral}

Table~\ref{tab:spectral} compares the equilibrium spectral radius of the interpolated family with that of the Grad closure for $M = 3, \ldots, 13$.
The Grad spectral radius equals the largest Gauss--Hermite quadrature node (the outermost zero of $\He_{M+1}$), which grows as $\sqrt{2(M+1)}$.
The interpolated family has a smaller spectral radius at every $M \geq 5$, allowing proportionally larger CFL time steps.
The CFL advantage grows from 13\% at $M = 5$ to 29\% at $M = 13$.
For $M = 3$, the two closures share the same equilibrium Jacobian.

\begin{table}
\caption{
Equilibrium spectral radius $\rho(A^0)$ for the interpolated family (all $\eta$) and the Grad Hermite expansion closure.
The CFL ratio is $\rho_{\mathrm{Grad}}/\rho_{\mathrm{RI}}$.
}
\label{tab:spectral}
\begin{ruledtabular}
\begin{tabular}{crrc}
$M$ & $\rho_{\RI}$ & $\rho_{\mathrm{Grad}}$ & CFL ratio \\
\colrule
 3 & 2.334 & 2.334 & 1.000 \\
 5 & 2.932 & 3.324 & 1.134 \\
 7 & 3.451 & 4.145 & 1.201 \\
 9 & 3.914 & 4.860 & 1.241 \\
11 & 4.337 & 5.501 & 1.268 \\
13 & 4.728 & 6.087 & 1.288 \\
\end{tabular}
\end{ruledtabular}
\end{table}

\section{PDE demonstrations}
\label{sec:pde}

The closed moment system~\eqref{eq:moment-hierarchy} is solved using a finite-volume method with Rusanov flux, fifth-order weighted essentially non-oscillatory (WENO-Z) reconstruction, and third-order strong-stability-preserving (SSP) Runge--Kutta time integration.
The numerical method is described in Appendix~\ref{app:solver}.
All runs use 800 cells, Courant--Friedrichs--Lewy (CFL) number $0.4$, and transmissive boundary conditions.

\subsection{Free-transport Riemann problem}
\label{sec:riemann-free}

The initial data follow the test case of Fox and Laurent~\cite{FoxLaurent2022}:
\begin{equation}
    (\rho, v, \theta)_L = (1, 1, \tfrac{1}{3}), \qquad
    (\rho, v, \theta)_R = (1, -1, \tfrac{1}{3}),
    \label{eq:riemann-free-ic}
\end{equation}
on $[-0.5, 0.5]$ with final time $T = 0.1$ and source $S = 0$.
The exact kinetic solution is known:
$f(t,x,c) = f_L(c)$ if $c > x/t$ and $f_R(c)$ if $c < x/t$, with moments computed by velocity-space quadrature with 10\,000 points.

\begin{figure}
\centering
\includegraphics[width=\columnwidth]{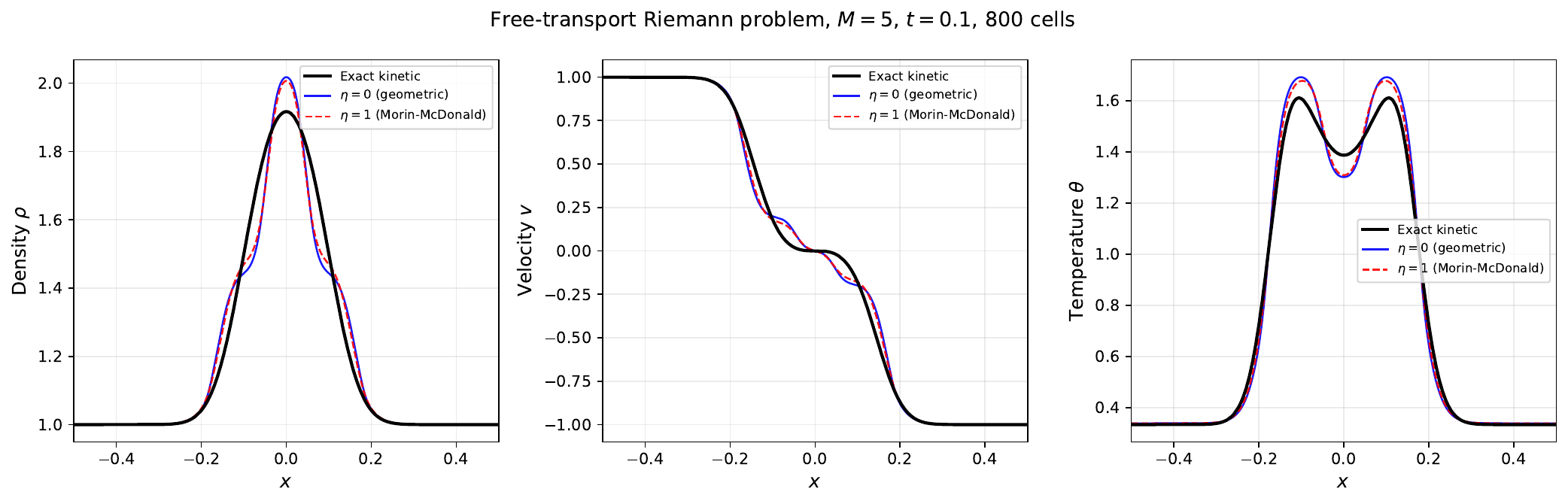}
\caption{Free-transport Riemann problem at $M = 5$, $t = 0.1$, 800 cells. Both the geometric ($\eta = 0$, solid blue) and Morin--McDonald ($\eta = 1$, dashed red) closures track the exact kinetic solution (black). Zero complex eigenvalue events and zero realizability failures in both runs.}
\label{fig:riemann-free}
\end{figure}

Figure~\ref{fig:riemann-free} compares the density, velocity, and temperature profiles at $t = 0.1$ for $\eta = 0$ and $\eta = 1$ against the exact kinetic reference at $M = 5$.
Both closures produce stable, sharp solutions with no spurious oscillations.
Table~\ref{tab:pde-errors} gives the $L^2$ errors for $M = 5, 7, 9, 11$.
At $M = 5$, $\eta = 1$ (MM) has slightly lower $L^2$ errors.
At $M = 7$, $9$, and $11$, $\eta = 0$ (geometric) is slightly more accurate in density and velocity.
In all cases, the difference between closures is smaller than the truncation error from one moment order to the next.
At $M = 9$, the density and velocity $L^2$ errors decrease relative to $M = 7$, but the temperature $L^2$ error increases.
A grid refinement study at 1600 cells confirms that this non-monotone behavior is not a spatial resolution artifact: the temperature $L^2$ error at 1600 cells ($5.05 \times 10^{-2}$) is essentially unchanged from 800 cells ($5.04 \times 10^{-2}$).
The spike is specific to $M = 9$: at $M = 11$, the temperature $L^2$ error drops back to $3.98 \times 10^{-2}$ (Table~\ref{tab:pde-errors}).
Density and velocity errors decrease monotonically across all tested moment orders.

\begin{table}
\caption{$L^2$ errors on the free-transport Riemann problem.  All runs: 800 cells, zero complex eigenvalue events, zero realizability failures.  The $M = 9$ temperature $L^2$ error is larger than $M = 7$. A 1600-cell grid refinement confirms this is a closure effect (not a spatial resolution artifact), and the error recovers at $M = 11$ (see text).}
\label{tab:pde-errors}
\begin{ruledtabular}
\begin{tabular}{cccccc}
$M$ & $\eta$ & $L^2(\rho)$ & $L^2(v)$ & $L^2(\theta)$ & Steps \\
\colrule
5 & 0 & $5.90 \times 10^{-2}$ & $5.21 \times 10^{-2}$ & $5.33 \times 10^{-2}$ & 1933 \\
5 & 1 & $5.18 \times 10^{-2}$ & $4.34 \times 10^{-2}$ & $4.22 \times 10^{-2}$ & 1995 \\
\colrule
7 & 0 & $3.10 \times 10^{-2}$ & $3.40 \times 10^{-2}$ & $3.27 \times 10^{-2}$ & 2199 \\
7 & 1 & $3.34 \times 10^{-2}$ & $3.68 \times 10^{-2}$ & $3.36 \times 10^{-2}$ & 2315 \\
\colrule
9 & 0 & $2.23 \times 10^{-2}$ & $2.07 \times 10^{-2}$ & $5.04 \times 10^{-2}$ & 2480 \\
9 & 1 & $2.07 \times 10^{-2}$ & $2.01 \times 10^{-2}$ & $4.99 \times 10^{-2}$ & 2640 \\
\colrule
11 & 0 & $2.05 \times 10^{-2}$ & $2.05 \times 10^{-2}$ & $3.98 \times 10^{-2}$ & 2704 \\
11 & 1 & $2.12 \times 10^{-2}$ & $2.08 \times 10^{-2}$ & $4.02 \times 10^{-2}$ & 2904 \\
\end{tabular}
\end{ruledtabular}
\end{table}

\begin{figure*}
\centering
\includegraphics[width=\textwidth]{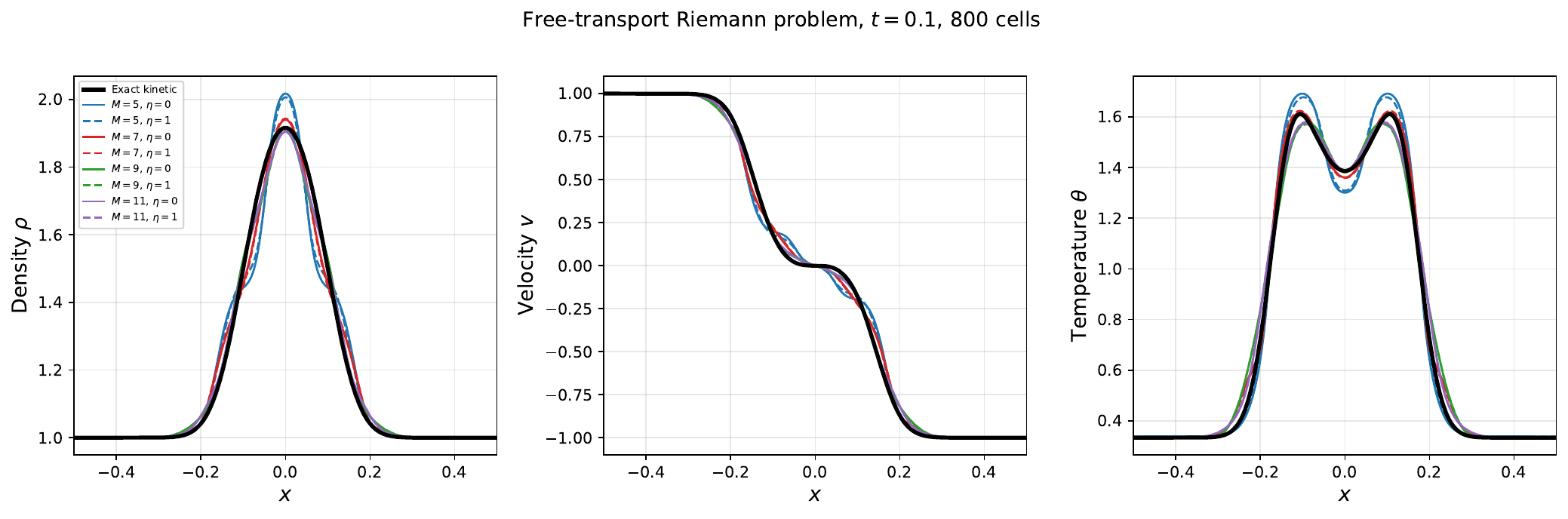}
\caption{Free-transport Riemann problem at $t = 0.1$, 800 cells, for $M = 5$ (blue), $M = 7$ (red), $M = 9$ (green), and $M = 11$ (purple), with both $\eta = 0$ (solid) and $\eta = 1$ (dashed), against the exact kinetic solution (black). Density and velocity converge monotonically with $M$. The $M = 9$ temperature profile shows a non-monotone $L^2$ error spike that recovers at $M = 11$ (Table~\ref{tab:pde-errors}).}
\label{fig:convergence-pde}
\end{figure*}

Figure~\ref{fig:convergence-pde} shows the $M = 5$, $7$, $9$, and $11$ solutions together.
Each increase in $M$ brings the density and velocity profiles closer to the exact kinetic reference.
The $M = 9$ temperature profile exhibits a non-monotone $L^2$ error spike that recovers at $M = 11$, consistent with the closure-level nature of this effect (Sec.~\ref{sec:discussion}).

\subsection{BGK Riemann problem}
\label{sec:riemann-bgk}

The same initial data~\eqref{eq:riemann-free-ic} are run with the BGK collision operator at two relaxation times: $\tau = 0.1$ (transition regime, $T/\tau = 1$) and $\tau = 0.01$ (near-continuum, $T/\tau = 10$).
All four runs ($\eta = 0, 1$ at both $\tau$) complete with zero complex eigenvalue events and zero realizability failures.

The maximum pointwise difference between $\eta = 0$ and $\eta = 1$ in the macroscopic fields is $3.3 \times 10^{-2}$ (free transport), $1.8 \times 10^{-2}$ ($\tau = 0.1$), and $1.1 \times 10^{-3}$ ($\tau = 0.01$).
The closure difference thus shrinks monotonically with decreasing Knudsen number: in the continuum limit, all closures produce the same macroscopic solution because the BGK source drives the distribution toward equilibrium, where the closure is exact.

\begin{figure*}
\centering
\includegraphics[width=\textwidth]{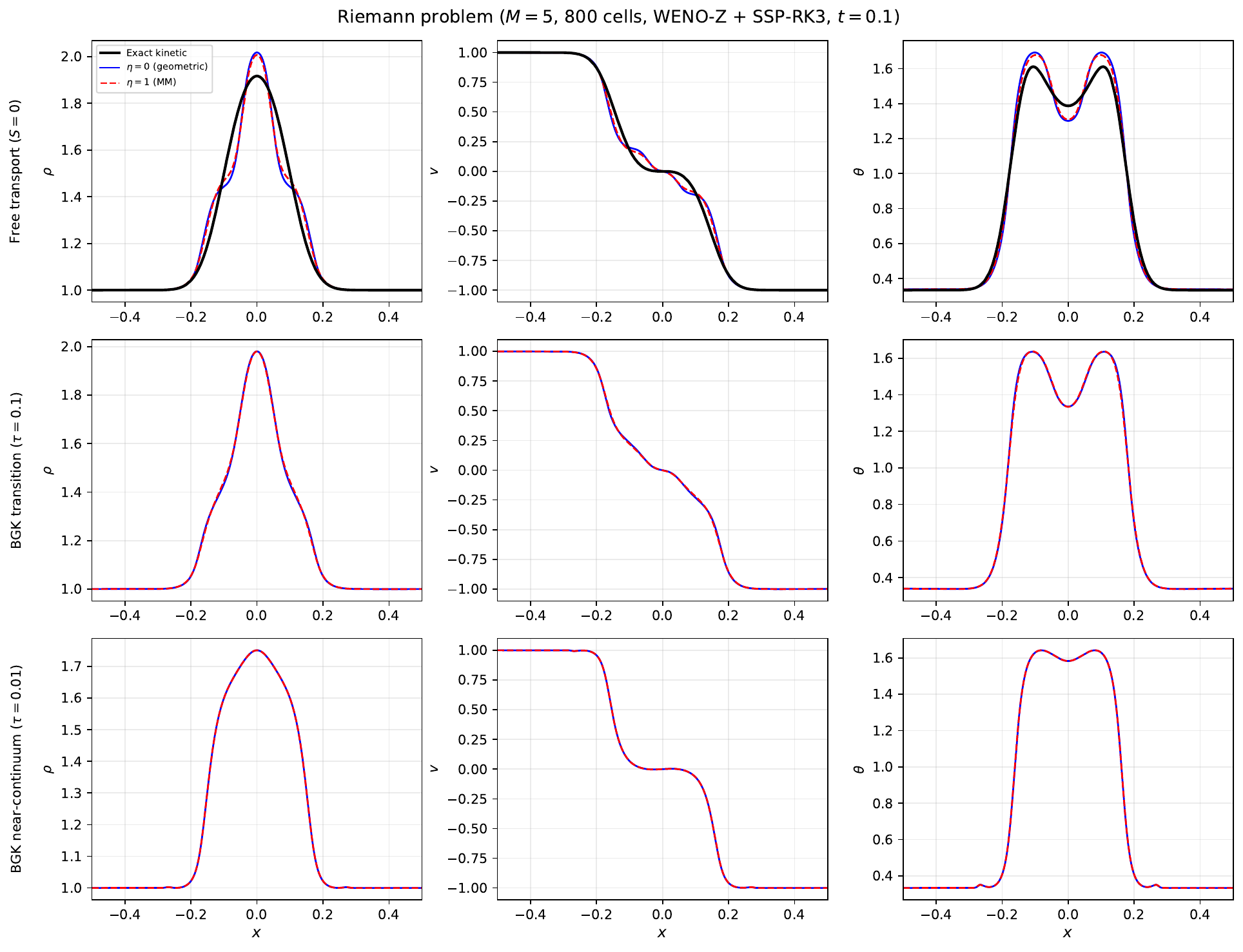}
\caption{Riemann problem at $M = 5$, $t = 0.1$, 800 cells.  Top: free transport ($S = 0$).  Middle: transition BGK ($\tau = 0.1$).  Bottom: near-continuum BGK ($\tau = 0.01$).  The closure difference between $\eta = 0$ (solid blue) and $\eta = 1$ (dashed red) shrinks monotonically with decreasing relaxation time. All runs: zero complex eigenvalue events, zero realizability failures.}
\label{fig:three-regimes}
\end{figure*}

Figure~\ref{fig:three-regimes} shows the density, velocity, and temperature profiles for the three regimes: free transport (top), transition (middle), and near-continuum (bottom).
The transition case shows a smoothed version of the free-transport solution, while the near-continuum case is nearly indistinguishable between $\eta = 0$ and $\eta = 1$.

At $M = 9$, the BGK Riemann problem at $\tau = 0.01$ also completes with zero complex eigenvalue events and zero realizability failures for both $\eta = 0$ (3194 steps) and $\eta = 1$ (3360 steps).
The maximum pointwise closure difference at $M = 9$ is $1.3 \times 10^{-4}$ (density), $2.1 \times 10^{-4}$ (velocity), and $3.4 \times 10^{-4}$ (temperature), roughly an order of magnitude smaller than at $M = 5$.
This confirms that higher moment order reduces the closure sensitivity in the near-continuum regime.

\section{Discussion}
\label{sec:discussion}

The decomposition $C = u_{2n}^{\min} + D$ places every odd moment closure in a common geometric framework.
The boundary term $u_{2n}^{\min}$ is the Schur complement of the Hankel matrix, the smallest admissible even moment.
The margin $D$ measures how far into the interior of moment space the closure places the extended moment vector.

The odd-order Gramian closure sets $D = 0$, saturating the boundary.
The constant-margin closure sets $D = \rho\, n!\, \theta^n$, using the Gaussian equilibrium value as a uniform estimate of the margin.
The geometric Schur-ratio closure scales the margin by the geometric mean of the normalized lower-order Schur depths.
The Morin--McDonald closure scales it by the arithmetic mean plus an $a$-dispersion term.
Each is a specific instance of the general form~\eqref{eq:ri-general}.

The accuracy-robustness tradeoff between the geometric and arithmetic endpoints has a precise origin.
The geometric mean produces the smallest margin by the AM--GM inequality, so it tracks the reference moment more closely and over-predicts less.
The arithmetic mean produces a larger margin, which leaves more room for real eigenvalues in the characteristic polynomial but over-predicts the moment.
A second-order expansion in the log normalized Schur-complement ratios $\ell_k = \log q_k$ gives $A_n = G_n[1 + \tfrac{1}{2}\sum w_k(\ell_k - \bar\ell)^2 + O(|\ell|^3)]$, where $\bar\ell = \sum w_k \ell_k$.
The endpoints differ only at second order in the Schur-ratio spread, which is why the accuracy difference is small (2--4\%), the equilibrium linearization is identical, and the error ratio $e(\eta{=}0)/e(\eta{=}1)$ converges to 1 at high moment orders (Table~\ref{tab:convergence}).

Global hyperbolicity for any odd-$M$ closure beyond $M = 3$ remains an open problem.
No existing closure, including the Morin--McDonald closure, has an analytic hyperbolicity proof at arbitrary realizable states.
The random stress test (Table~\ref{tab:random}) shows that the geometric endpoint has a small but nonzero failure rate at extreme states ($R_2$ far from 1), while the Morin--McDonald endpoint is more robust, with failures observed only at near-boundary states with Hankel condition numbers above $10^{12}$.
The family can be extended to two parameters $(\eta, \beta)$ by writing $\alpha_{\eta,\beta,n} = R_{n-1}[(1-\eta)G_n + \eta A_n + \beta S_n]$.
At equilibrium, $S_n = 0$, so the equilibrium Jacobian is unchanged for any $\beta$.
The current family has $\beta = 1$.
Increasing $\beta$ strengthens the coupling to the top odd moment through the recurrence-coefficient differences, which could improve robustness on states where the Schur-ratio margin alone is insufficient.

The PDE tests in Sec.~\ref{sec:pde} show that the closed moment system produces stable, sharp solutions on the free-transport Riemann problem at $M = 5$, $7$, $9$, and $11$ and on the BGK Riemann problem at $M = 5$ and $9$.
Neither endpoint of the family exhibits complex eigenvalues or realizability violations in any of the computed cases.
At $M = 5$, the arithmetic endpoint is slightly more accurate in $L^2$ norm on the free-transport test, while at $M = 7$, $9$, and $11$ the geometric endpoint is more accurate.
The crossover reflects the interplay between closure error and spatial discretization.
At $M = 5$, the spatial error (${\sim}2 \times 10^{-2}$) is comparable to the closure difference and partially masks the geometric endpoint's advantage.
At $M \geq 7$, the closure error is dominant and the static accuracy advantage of $\eta = 0$ (Table~\ref{tab:bimodal}) becomes visible.
In the static benchmark, the relative advantage of $\eta = 0$ shrinks with moment order (Table~\ref{tab:convergence}), because the AM--GM ratio $A_n/G_n$ approaches 1 as higher $M$ resolves more of the distribution.
The closure difference shrinks monotonically with decreasing $\tau$: at $\tau = 0.01$ ($T/\tau = 10$), the two endpoints differ by at most 0.1\% in any macroscopic field.
For practical use, the Morin--McDonald closure ($\eta = 1$) remains the recommended choice: it has the largest margin, no failures at $M = 5$ and $7$ and only a single near-boundary failure at $M = 9$ (shared by all members of the family), and its 2--4\% accuracy cost relative to the geometric endpoint is negligible compared to the truncation error of the moment hierarchy itself.
The contribution of the present work is structural: it explains \textit{why} the Morin--McDonald closure works (it is the arithmetic Schur-ratio margin) and \textit{why} simpler constructions cannot (the macroscopic fields alone do not determine the margin).

The non-monotone temperature $L^2$ error at $M = 9$ (Table~\ref{tab:pde-errors}) is a truncation-level effect, not a closure-level one: both endpoints $\eta = 0$ and $\eta = 1$ produce essentially the same temperature spike ($5.04 \times 10^{-2}$ and $4.99 \times 10^{-2}$ respectively), and a 1600-cell grid refinement leaves it unchanged ($5.05 \times 10^{-2}$).
The spike is localized near $x/t \approx \pm 2$, where the exact distribution transitions from strongly bimodal (near $x = 0$) to nearly unimodal (near $x = \pm 0.5$).
In the free-transport Riemann problem, the distribution has a velocity-space discontinuity at $c = x/t$, and the $M = 2n{-}1$ moment system must represent this discontinuity using $n$ orthogonal polynomial roots.
Non-monotone convergence in the $L^2$ temperature error is analogous to non-monotone convergence in polynomial approximation of non-smooth functions: the specific pattern of $n = 5$ characteristic velocities at $M = 9$ produces a less favorable representation of the temperature profile in the transition region than either $n = 4$ ($M = 7$) or $n = 6$ ($M = 11$).
The density and velocity, which are lower-order moments less sensitive to the tail structure of the distribution, converge monotonically across all $M$.
A full analysis of this non-monotonicity requires characterizing how the roots of $p_n$ interact with the velocity-space support of the truncated distribution at each spatial location.
This is an open problem for truncated moment methods in general, not specific to the closure construction proposed here.

The Grad Hermite expansion closure~\cite{Grad1949} performs poorly on both benchmarks: its bimodal errors are 15--20\% and increase with $M$ (Table~\ref{tab:comparison}), and it loses strict hyperbolicity on 31--45\% of bimodal states.
At equilibrium, the Grad flux Jacobian has eigenvalues at the Gauss--Hermite quadrature nodes (zeros of $\He_{M+1}$), giving spectral radii that grow as $\sqrt{2(M{+}1)}$, while the interpolated family has smaller spectral radii at every $M \geq 5$ (Table~\ref{tab:spectral}).
The CFL advantage grows from $13\%$ at $M = 5$ to $29\%$ at $M = 13$.
For $M = 3$, both closures share the same equilibrium Jacobian, but for $M \geq 5$ the Jacobians differ.
The CFL advantage is a direct consequence of the Jacobian universality (Sec.~\ref{sec:family}): all closures sharing the family's equilibrium gradient $\nabla\gamma\big|_G$ inherit the same (smaller) spectral radius, while the Grad closure's Hermite expansion produces eigenvalues at the Gauss--Hermite quadrature nodes.
The PDE tests in Sec.~\ref{sec:pde} compare only the two endpoints of the interpolated family ($\eta = 0$ and $\eta = 1$).
The Grad Hermite expansion closure is not included because our solver does not implement it.
The Grad closure used in the static benchmarks (Tables~\ref{tab:comparison}--\ref{tab:spectral}) is the full Hermite expansion, which depends on all moments $u_0, \ldots, u_M$ and is real-diagonalizable at equilibrium.
This should not be confused with the simplified Grad closure $C = g_{2n}(\rho, v, \theta)$, which depends only on density, velocity, and temperature, and whose non-hyperbolicity at equilibrium for $M \geq 5$ is proved in Sec.~\ref{sec:structural}.

The entropy analysis in Sec.~\ref{sec:entropy} provides the first entropy structure results for any Hankel-based odd-$M$ closure.
The linearized entropy exists at every tested order, and BGK dissipation holds for a source-compatible choice of the symmetrizer weights (verified for $M = 3, 5, 7, 9$), establishing near-equilibrium thermodynamic consistency.
The nonexistence of a smooth nonlinear entropy, certified for $M = 5$ and $M = 7$, is independent of $\eta$ and the equilibrium state, suggesting a structural obstruction shared by the entire interpolated family and the Morin--McDonald closure.
For $M = 3$, the smooth entropy exists, but its physical interpretation and connection to the kinetic entropy $-\int f \log f\, \mathrm{d}c$ remain open.
The Morin--McDonald paper~\cite{MorinMcDonald2025} does not contain an entropy analysis, and the present result fills this gap.

Full gauge invariance (affine covariance) of the interpolated family has not been established beyond the equilibrium Jacobian matching derived in Sec.~\ref{sec:schur-ratios}.
Multi-dimensional extensions are left for future work.

\section{Conclusions}
\label{sec:conclusions}

We have derived the geometric structure of odd-order moment closures for the one-dimensional kinetic equation.
The results, with their scope, are as follows.

The squared norm of the Gaussian orthogonal polynomial, which equals the difference $g_{2n} - g_{2n}^{\min}$ between the Gaussian moment and its Schur-complement lower bound, is exactly $\rho\, n!\, \theta^n$ (Eq.~\eqref{eq:eq-gap}, derived from the squared norm of the monic Hermite polynomial, verified symbolically for $n = 2, 3, 4$).
The characteristic polynomial of any closure in the interior of moment space is exactly $p_n^2 - \mathcal{D}$, where $p_n$ is the monic orthogonal polynomial and $\mathcal{D}$ is the derivative polynomial of the margin (Eq.~\eqref{eq:structural-identity}).
The equilibrium derivatives of any $(\rho, v, \theta)$-only margin are uniquely fixed by dimensional analysis~\eqref{eq:covariance}, so the equilibrium characteristic polynomial is the same for every such margin.
This polynomial has complex roots for $n = 3$ (proved exactly by Vieta's formula) and for $n = 4, 5$ (verified numerically).

The interpolated family~\eqref{eq:closure-family} parametrizes closures by a scalar $\eta \in [0,1]$.
Every member preserves Gaussian equilibrium and shares the same equilibrium Jacobian.
The identity $C_{1,n} = C_{\MM,n}$ is exact, giving the Morin--McDonald closure a geometric interpretation as the arithmetic Schur-ratio margin above the realizability boundary.
The ordering $D_0 \leq D_\eta \leq D_1$ follows from the weighted AM--GM inequality and explains why the Morin--McDonald closure ($\eta = 1$), as the arithmetic endpoint, provides the largest margin and the most robust among all tested members (1000 random states per $M$, complex-step Jacobians).
The geometric endpoint ($\eta = 0$) provides the smallest margin and is 2--4\% more accurate on the bimodal benchmark (71 states per $M$, $M = 5, 7, 9$) and 8\% more accurate on the Mott--Smith shock benchmark ($\mathrm{Ma} = 4$), at the cost of a small failure rate on random states (1\% at $M = 5$, decreasing with $M$).

The Jacobian universality theorem shows that the equilibrium Jacobian depends on the closure only through $\nabla\gamma\big|_G$: the interpolated family and Morin--McDonald share a common equilibrium gradient and therefore a common Jacobian, whose spectral radii are $13\%$ ($M = 5$) to $29\%$ ($M = 13$) smaller than those of Grad's closure (Table~\ref{tab:spectral}).
Comparison with the Gramian and Grad closures (Table~\ref{tab:comparison}) shows that the interpolated family achieves errors 10--40$\times$ smaller than the Gramian and the Grad Hermite expansion, while maintaining zero hyperbolicity failures on the bimodal benchmark (the Grad closure fails on 31--45\% of states).

A linearized entropy with Vandermonde-diagonal symmetrizer exists at every tested order (through $M = 17$).
For a source-compatible choice of the symmetrizer weights, the BGK source dissipates it near equilibrium (verified for $M = 3, 5, 7, 9$).
A smooth nonlinear entropy exists for $M = 3$ (verified by linear programming feasibility) but does not exist for $M = 5$ or $M = 7$ (certified by LP infeasibility, independent of $\eta$ and the equilibrium state).
This is the first entropy analysis for any Hankel-based odd-$M$ closure.

PDE simulations of the free-transport Riemann problem at $M = 5$, $7$, $9$, and $11$ and the BGK Riemann problem at $M = 5$ and $9$ produce stable solutions with zero hyperbolicity failures and zero realizability violations.
The closure difference between $\eta = 0$ and $\eta = 1$ is at most 3\% of the macroscopic fields in the free-transport regime and shrinks monotonically with decreasing relaxation time.
Density and velocity $L^2$ errors decrease monotonically from $M = 5$ to $M = 11$.
The temperature $L^2$ error has a non-monotone spike at $M = 9$ (confirmed by grid refinement to be a closure effect, not a spatial resolution artifact) that recovers at $M = 11$ (Table~\ref{tab:pde-errors}).

Global hyperbolicity of odd-$M$ closures remains unproven for any closure.
The geometric framework developed here may provide a path toward such a proof, since the characteristic polynomial has the explicit form $p_n^2 - \mathcal{D}$ and the margin $D$ is a smooth, positive, and explicitly constructed function of the retained moments.

\appendix

\section{Explicit closure formulas}
\label{app:explicit}

For $M = 5$ ($n = 3$), the Schur data are $H_2 = (u_{i+j})_{i,j=0}^{2}$ and $b = (u_3, u_4, u_5)^T$.
The boundary value is $u_6^{\min} = b^T H_2^{-1} b$.
The normalized Schur ratio is $R_2 = \sigma_{2,2}/(2\rho\theta^2)$, where $\sigma_{2,2} = u_4 - (u_2, u_3)\, H_1^{-1} (u_2, u_3)^T$ and $H_1 = \bigl(\begin{smallmatrix} u_0 & u_1 \\ u_1 & u_2 \end{smallmatrix}\bigr)$.
The closure is
\begin{equation}
    C_{\eta,3} = u_6^{\min} + 6\rho\theta^3 R_2\!\left[(1-\eta)R_2^{2/3} + \eta\!\left(\tfrac{1}{3} + \tfrac{2R_2}{3}\right) + S_3\right],
\end{equation}
where $S_3 = [(a_2-a_0)^2 + (a_2-a_1)^2]/(6\theta)$ and $a_0, a_1, a_2$ are the Chebyshev recurrence coefficients computed from $u_0, \ldots, u_5$.

For $M = 7$ ($n = 4$), $R_2$ and $R_3 = \sigma_{3,3}/(6\rho\theta^3)$ are both needed.
The closure is
\begin{equation}
    C_{\eta,4} = u_8^{\min} + 24\rho\theta^4 R_3\!\left[(1-\eta)R_2^{-1/6}R_3^{1/2} + \eta\!\left(\tfrac{1}{6} + \tfrac{R_2}{3} + \tfrac{R_3}{2R_2}\right) + S_4\right],
\end{equation}
where $S_4 = \sum_{\ell=0}^{3}(a_3 - a_\ell)^2/(12\theta)$.

\section{Equilibrium characteristic polynomial coefficients}
\label{app:polynomials}

At the standard Gaussian equilibrium $(\rho, v, \theta) = (1, 0, 1)$, the closed system with the interpolated closure has the same characteristic polynomial for all $\eta$.
By symmetry, only even powers of $\lambda$ survive.
Substituting $\mu = \lambda^2$, the equilibrium polynomial $q_n(\mu) = \mu^n + a_{n-1}\mu^{n-1} + \cdots + a_0$ has integer coefficients listed in Table~\ref{tab:poly}.

The subleading coefficient satisfies $a_{n-1} = -(n^2 + 2)$ for $n = 2, \ldots, 9$ (observed numerically to integer precision, no proof).
All $\mu$-roots are positive for every $n$ tested, confirming hyperbolicity at equilibrium through $M = 17$ (numerical).

\begin{table}
\caption{
Coefficients of the equilibrium $\mu$-polynomial $q_n(\mu)$ for $n = 2, \ldots, 7$.
The polynomial has degree $n$ and leading coefficient 1.
}
\label{tab:poly}
\begin{ruledtabular}
\begin{tabular}{cl}
$n$ & $q_n(\mu)$ \\
\colrule
2 & $\mu^2 - 6\mu + 3$ \\
3 & $\mu^3 - 11\mu^2 + 21\mu - 3$ \\
4 & $\mu^4 - 18\mu^3 + 80\mu^2 - 90\mu + 15$ \\
5 & $\mu^5 - 27\mu^4 + 216\mu^3 - 600\mu^2 + 495\mu - 45$ \\
6 & $\mu^6 - 38\mu^5 + 477\mu^4 - 2436\mu^3 + 4935\mu^2 - 3150\mu + 315$ \\
7 & $\mu^7 - 51\mu^6 + 923\mu^5 - 7485\mu^4 + 28035\mu^3 - 44625\mu^2$ \\
  & $\quad + 23625\mu - 1575$ \\
\end{tabular}
\end{ruledtabular}
\end{table}

\section{Complex-step differentiation}
\label{app:complex-step}

The Jacobian last row $C_{u_j}$ is computed by complex-step differentiation~\cite{Squire1998}:
\begin{equation}
    \frac{\partial C}{\partial u_j} = \frac{\mathrm{Im}[C(u + ih\, e_j)]}{h}, \qquad h = 10^{-30}.
\end{equation}
This requires the closure to accept complex-valued moment vectors.
All operations in the closure (Hankel inversion, logarithms, exponentials, Chebyshev recurrence) support complex arithmetic in standard numerical libraries.
The method has no subtractive cancellation error, so it gives machine-precision derivatives regardless of the Hankel condition number.
This is confirmed by comparing with analytic derivatives at equilibrium, where the discrepancy is below $10^{-12}$ for all $M$ tested.

\section{Numerical solver}
\label{app:solver}

The moment system~\eqref{eq:moment-hierarchy} is discretized on a uniform mesh with cell width $\Delta x$ and cell-centered unknowns $U_i = (u_{0,i}, \ldots, u_{M,i})^T$.
The numerical flux at each interface $i+\tfrac{1}{2}$ is the Rusanov flux
\begin{equation}
    F_{i+1/2} = \tfrac{1}{2}(F_L + F_R) - \tfrac{1}{2}\,a_{\max}(U_R - U_L),
\end{equation}
where $F_L$, $F_R$ are the physical fluxes evaluated at left and right reconstructed states, and $a_{\max} = \max(\rho(A_L), \rho(A_R))$ is the maximum spectral radius of the flux Jacobian.

Left and right states are obtained by fifth-order WENO-Z reconstruction~\cite{BorgesEtAl2008} applied independently to each conserved variable $u_k$.
The spectral radius $\rho(A)$ is computed at each interface by forming the companion-type flux Jacobian
\begin{equation}
    A =
    \begin{pmatrix}
        0 & 1 &  &  \\
          & 0 & \ddots &  \\
          &   & \ddots & 1 \\
        \partial C/\partial u_0 & \partial C/\partial u_1 & \cdots & \partial C/\partial u_M
    \end{pmatrix}
\end{equation}
and calling LAPACK \texttt{dgeev} for the eigenvalues.
The closure derivatives $\partial C/\partial u_j$ are computed by a tangent-linear pass through the closure evaluation: the Chebyshev forward algorithm, Schur-complement solves, and geometric/arithmetic mean computations are differentiated analytically, producing exact last-row entries for the companion matrix at $O(n^3)$ cost per cell (comparable to the forward evaluation itself).

Time integration uses the third-order strong-stability-preserving Runge--Kutta scheme~\cite{ShuOsher1988}:
\begin{align}
    U^{(1)} &= U^n + \Delta t\, L(U^n), \notag \\
    U^{(2)} &= \tfrac{3}{4}\,U^n + \tfrac{1}{4}\bigl(U^{(1)} + \Delta t\, L(U^{(1)})\bigr), \\
    U^{n+1} &= \tfrac{1}{3}\,U^n + \tfrac{2}{3}\bigl(U^{(2)} + \Delta t\, L(U^{(2)})\bigr), \notag
\end{align}
where $L$ includes both the flux divergence and the BGK source term.
When a BGK source is present with relaxation time $\tau$, it is evaluated at each stage:
$S_k = (g_k - u_k)/\tau$ for $k \geq 3$ and $S_k = 0$ for $k = 0, 1, 2$, where $g_k$ are the Gaussian moments with the same $(\rho, v, \theta)$.
The time step is $\Delta t = \mathrm{CFL}\,\Delta x / a_{\max}$, further limited to $\Delta t \leq 0.1\,\tau$ when a BGK source is active.

At each Runge--Kutta stage, the updated state is checked for realizability: $\rho > 0$, $\theta > 0$, and $\sigma_{j,j} > 0$ for $j = 1, \ldots, n-1$.
If a reconstructed state at an interface fails the realizability check, it is replaced by the cell-average state.
In all runs reported in this paper, no cell-average state ever fails realizability.

The Rusanov flux is chosen over more sophisticated Riemann solvers (Roe, HLLC) because the moment system has no closed-form eigenvectors for $M \geq 5$, making field-by-field decomposition impractical.
The Rusanov flux requires only the spectral radius and is unconditionally stable for any hyperbolic system.
Fifth-order WENO-Z provides sufficient resolution of discontinuities without the cost of characteristic decomposition.
Componentwise reconstruction on the conserved moments is standard for moment systems where the eigenvectors are unavailable analytically~\cite{FoxLaurent2022}.
Third-order SSP-RK matches the formal accuracy of the spatial discretization after the WENO limiter (which reduces to third order near discontinuities) while preserving the TVD property of the underlying first-order scheme.

\begin{acknowledgments}
The author acknowledges the financial support provided by the National Science and Technology Council (NSTC) under Project No.\ NSTC 113-2628-E-006-005-MY3.
\end{acknowledgments}

\section*{Author Declarations}
\subsection*{Conflict of Interest}
The author has no conflicts to disclose.

\subsection*{Data Availability}
The data that support the findings of this study are available within the article. The code used for numerical validation in this study is available from the author upon reasonable request.


\end{document}